\documentclass[prb,aps,epsfig,twocolumn,showpacs]{revtex4-1}
\usepackage{color}
\usepackage{cancel}
\usepackage{lipsum}
\usepackage{epsfig,amssymb,amsmath,latexsym}
\graphicspath{{chobi/}}
\usepackage{float}
\usepackage{color}
\usepackage[colorlinks=true,linktoc=page,linkcolor=magenta,citecolor=magenta]{hyperref}

\begin{document}
\textheight=23.8cm

\title{Dynamical spin-spin  susceptibility of Silicene}
\author{ Surajit Sarkar, Suhas Gangadharaiah}
\affiliation {Department of Physics, Indian Institute of Science Education and Research, Bhopal, India}

\date{\today}

\begin{abstract}

We present a detailed study of the  imaginary and real parts of the spin-susceptibility  of silicene which can be generalized to other buckled honeycomb structure. We find   that while the off-diagonal components are non-zero in individual valleys,  they add up to zero upon including contributions from both the valleys. We investigate  the interplay of the spin-orbit interaction and an external electric field applied perpendicular to the substrate and  find that      although the  $xx$ and $yy$ components  of  the susceptibility are identical,  they differ from the  $zz$-component.
The  external electric field  plays an important role in 
modifying the allowed inter-subband regions.  
In the dynamic limit, the real part of the susceptibility  exhibits log-divergence, position of which can be tuned  by the electric field  and therefore has implications for spin-collective excitations. The effect of the electric field  on the  static part of the susceptibility and its consequence for the long distance decay of the spin-susceptibility  have been explored. 
	
\end{abstract}

\maketitle


\section{ Introduction} {\label{sec:I}}
Spin-orbit (SO) interaction is one of the key ingredients in a spintronics device required  for controlling  and manipulating  the spin degrees of freedom of an electron via the electric field~\cite{haus1955,datta1990}. In this regard,  enormous progress has already been made in the study of semiconductor based spintronics device~\cite{haus1955,das2004}.  Recently the possibility of graphene and other 2D materials, in particular, silicene and germanene~\cite{yao2011},  topological insulators~\cite{raghu2010,pietro2013}, Weyl semimetals~\cite{das2015,pesin2014,xiao2015} along with monolayer transition metal dichalcogenides such as MoS$_2$~\cite{wang2012} with intrinsic and extrinsic SO coupling, have garnered wide attention from the fundamental physics point of view as well as for their potential  for spintronics applications.

The low energy effective theory of many of these new materials is governed by the Dirac physics. In graphene, due to the  relatively small mass of carbon atoms the SO coupling is very weak therefore the physics is  effectively described by the massless Dirac theory. The conduction and valance bands meet at the two inequivalent Dirac points, called the $K$ and $K'$ points, which is where the Fermi energy also lies. On the other hand, due to the higher mass of silicon atoms SO coupling in silicene is appreciable ($\approx$ 3.9meV)~\cite{motohiko1,vogta2012,lalmi2010}. 
Unlike graphene which is completely planar, silicene has a  buckled honeycomb sublattice structure resulting in the explicit breaking of inversion symmetry~\cite{liu2011}. An electric field applied perpendicular to the silicene surface leads to a staggered potential which in combination with the SO term determines the gap in the energy spectrum. 
Consequently, electric field can be used as a control parameter to drive  silicene from a trivial band insulator phase to symmetry protected topological phase (e.g spin Hall insulator~\cite{kane2005,zhang2006}). At the critical point the band-gap closes~\cite{drummond2012,motohiko2012} and silicene enters into a valley-spin polarized metallic state~\cite{ezawa2012prl,nicol2014, tcn2015}. These features in the energy spectrum provide the possibility for detecting quantum, anomalous and valley hall effects in silicene~\cite{ezawa2012prl,ezawa2013,nicol2013}.

Useful insights into the electronic properties of  materials are obtained by studying their charge response function or the charge polarization operator. It yields  information regarding the single particle and  collective excitations which are   crucial for understanding the static and dynamical properties of many body systems~\cite{stern1967,stern1982}.
While the  modifications  to the response function due to the SO coupling in 2DEG with parabolic dispersion have been investigated in great detail~\cite{raikh1999,gritsev2006}, it is only recently   that similar  studies on the charge response function of materials with Dirac like dispersion have been made~\cite{shung1986,gorbar2002,rafael2013,pyat2009,andreas2012,yafis2007,hwang2007,neto2012,raghu2010,pietro2013,das2015,pesin2014,xiao2015}. 
There have also been  studies on the spin response of  SO coupled 2D electron system and  of the helical surface states of a 3D topological insulator~\cite{shekhter2005,chesi2010,maslov2012,maiti2015,raghu2010}. By considering the dynamical spin-susceptibility  of  SO coupled 2D electron system the existence of spin-collective excitations was established~\cite{maiti2015}, moreover, the surface states of a 3D topological insulator, described by the Dirac spectrum, have  been predicted to host hybridized spin-charge coupled plasmons~\cite{raghu2010}. 
Recently, Raman spectroscopy was used to reveal the collective spin-excitations of the chiral surface states of  the  three dimensional topological insulator $\text{Bi}_2\text{Se}_3$~\cite{maiti2017}. 
On the other hand, the modifications to  the static spin-susceptibility due to the SO terms yield  additional interaction terms  like Dzyaloshinskii-Moriya and Ising terms  besides the usual isotropic Rudermann-Kittel-Kasuya-Yosida (RKKY) interaction term~\cite{imamura2004, jelena2013,chang2015, duan2018}.

Recent studies of the charge polarization function of silicene have predicted the existence of charge collective excitation with a $\sqrt{q}$ dispersion at small $q$~\cite{nicol2014,yao2014,peeters2014}. However, the  study of spin collective modes in silicene and other buckled honeycomb lattice is an ongoing and challenging work. As a first step towards the better understanding of  the role of spin-orbit interaction in silicene we study in detail the spin-susceptibility in the non-interacting limit. The imaginary part of the spin-susceptibility, which yields information regarding the single particle spin decay channel, are identical  for the  $xx$ and $yy$ components while the $zz$-components are different.  We discuss in detail the allowed single-particle transitions and the regions in the $(q,\omega)$ plane where the imaginary part of the susceptibility is non-zero. The role of electric field in extending the allowed  regions for particle-hole excitations is examined. 
We calculate the real part of susceptibility, with particular emphasis on the dynamic and static limits. We show that the real part of spin-susceptibility exhibits log-divergence in the dynamic limit (in the $xx$ and $yy$ channels) and discuss its significance with regard to the spin-collective modes. 
The static part of the spin-susceptibility exhibits Kohn-anomaly, interestingly nature of this anomaly and the momentums at  which this happens  can be controlled by electric field.  The consequence of it for the long distance decay behavior of the spin-susceptibility have been studied. 

Our paper is organized as follows: In sec.\ref{sec:II} we provide a general description of our model along with the  low energy effective Hamiltonian of silicene. In sec.~\ref{sec:III} we define the spin-susceptibility operator  and discuss the contributions to the imaginary part of the spin-susceptibility arising from  different transition scenarios. In sec.~\ref{sec: Real_q_0}  the real part of the spin-susceptibility in the dynamical and  static limits have been calculated.  Summary of the results are provided in sec.~\ref{sec:V}.

\section{ Model} {\label{sec:II}}
The tight binding Hamiltonian  of 2D silicene is given by
\begin{eqnarray}\label{eq:Hamiltonian}
H=-t\sum\limits_{\langle i,j\rangle{\alpha}}\hat{c}^{\dagger}_{i{\alpha}}\,\hat{c}_{j{\alpha}}
+i {\frac{{\lambda_{SO}}}{3{\sqrt{3}}}} \sum_{\langle\langle i,j\rangle\rangle{\alpha}{\beta}}{\nu}_{ij}\hat{c}^{\dagger}_{i{\alpha}}{\sigma}^{z}_{{\alpha}{\beta}}\,\hat{c}_{j{\beta}}\notag\\
+l\sum_{i{\alpha}}{\zeta_{i}}E^i_{z}\hat{c}^{\dagger}_{i{\alpha}}\,\hat{c}_{i{\alpha}} \,\,-{\mu}\sum_{i{\alpha}}\hat{c}^{\dagger}_{i{\alpha}}\,\hat{c}_{i{\alpha}}\ , \qquad
\end{eqnarray}
where the first term represents the nearest-neighbor hopping on the honeycomb lattice, the second term represents the effective SO term which couples   next nearest-neighbor sites. The coupling parameter is  denoted by $\lambda_{SO}$,  $\sigma^z$ the pauli spin matrix, ${\nu}_{ij}=\hat{z}\cdot({\vec{d}_i}\times{\vec{d}_j})/|{{\vec{d}_i}	\times{\vec{d}_j}}|$ with $\vec{d}_i$ and ${\vec{d}_j}$ being the bonds between the two next nearest-neighbor sites. The third  term represents the staggered sublattice potential, where ${\zeta}_i=\pm1$ for the A(B) sites and $2l$ is the separation between the A and B sublattices in the z-direction, $E_z$ is an applied electric field perpendicular to the plane and $\mu$ is the  chemical potential. For silicene $t=1.6~\text{eV}$, $\lambda_{SO}=3.9~\text{meV}$ and  $l=0.23\AA$~\cite{ezawa2012,ezawa2012prl,liu2011}.
The Hamiltonian receives an additional  contribution due to the Rashba SO-term, however, the magnitude of this term ($\lambda_R=0.7~\text{meV}$)   is almost an order of magnitude  less than $\lambda_{SO}$. Moreover,
 near the Dirac points the rashba term is given by the linear  $\sim\lambda_R k$ term which can be neglected when describing the low-energy physics~\cite{ezawa2012,ezawa2012prl}.
 We note that germanene which  has a buckled structure is also described by the Hamiltonian given in Eq.~\ref{eq:Hamiltonian}, with $t=1.3~\text{eV}$, $\lambda_{SO}=43~\text{meV}$ and $l=0.33\AA$~\cite{ezawa2012,ezawa2012prl,liu2011,yao2011},  here also the Rashba term can be neglected when describing the low energy physics.

\begin{figure}
\centering
\includegraphics[width=0.9\linewidth]{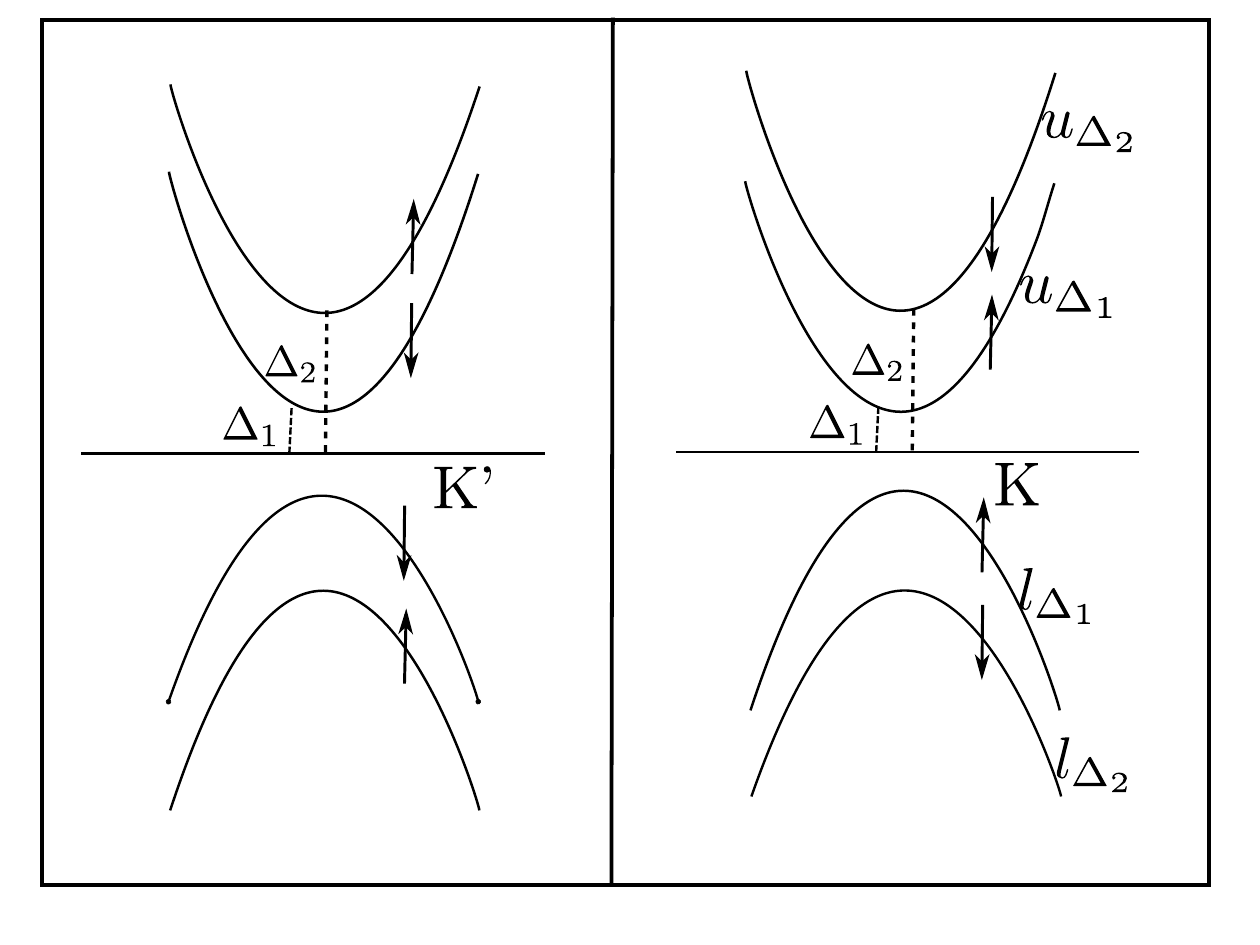}
\caption{(Color online) Energy spectrum near the $\text{K}$ and $\text{K}'$ points. The arrows indicate the orientation  of the spin in the respective band.}
\label{E-Spectrum}
\end{figure}

The low-energy effective Hamiltonian  about the two inequivalent Dirac points $K_{\eta}$ (where $\eta = \pm 1$) in the basis $({\psi_{A\uparrow}},{\psi_{B\uparrow}},{\psi_{A\downarrow}},{\psi_{B\downarrow}})$ 
acquires the form 
\begin{eqnarray}
&&H_\eta=\hbar v_F\Big(k_x(\hat{I} \otimes \hat{\tau}_1) -\eta k_y(\hat{I} \otimes \hat{\tau}_2)\Big) + lE_Z(\hat{I} \otimes \hat{\tau}_3) \nonumber\\
&&-\eta \lambda_{SO}\sigma^z\otimes \hat{\tau}_3\,\,\,\,\,\,\,\,\,\,\,\
\end{eqnarray}
where the Pauli-matrix $\hat{\tau}$ acts on the  sublattice basis and $\eta$ is the valley index. Henceforth, we will set 
$v_F=1$ and $\hbar=1$. In the presence of both  the electric field and the spin-orbit term the spectrum is given by
$\epsilon_{k_{\eta \beta}} = \alpha \sqrt{k^2 + \Delta_{\eta,\beta}^2 }$, where $\alpha=\pm 1$ and  the inequivalent gaps for spins $\beta=\pm 1$ are given by  $\Delta_{\eta,\beta}=|lE_z - \eta \beta \lambda_{SO}|$. In Fig.~\ref{E-Spectrum} we plot the energy spectrum near the $\text{K}$, $\text{K}'$ points, where the energy  gaps are  $\Delta_{1/2} = |lE_z \mp \lambda_{SO}| $. We note that the strength of the gap can be tuned by  external electric fields, in particular, for  the critical field $E_z^c= \lambda_{SO}/l$  the Hamiltonian exhibits gapless modes. 


\section{ Polarization  Function}{\label{sec:III}}
The non-interacting generalized susceptibility in the Matsubara formalism is given by~\cite{maiti2015}
\begin{eqnarray}
\chi_{ij} (q ,\omega_n) = -\int _P \text{Tr} \Big [\hat{\sigma}_i~ \hat{G}_P~\hat{\sigma}_j~  \hat{G}_{P+Q} \Big ],\label{eq:PF}
\end{eqnarray}  
where  $\text{Tr}$ denotes  trace over spin and sublattice degrees of freedom, $i,j = 0,x,y,z $, $P=(\vec{p},\Omega_n)$ and  $Q=(\vec{q},\omega_n)$. Note that the polarization function/operator
is related to the  susceptibility via the  relation, $\Pi_{ij} (q ,\omega_n) =-\chi_{ij}(q ,\omega_n)$. In the rest of the text we will be using the two terms interchangeably.

 The corresponding zero temperature Matsubara Green's function used in the above equation has the following form
\begin{eqnarray}
\hat{G_p}=\frac{1}{4}\sum_{\beta,\alpha =\pm 1} \frac{ \Bigg [ \big (\hat{I} + \beta \hat{\sigma}_3 \big )  \otimes \big (\hat{I}-\alpha (\vec{p}_\beta \cdot \vec{\tau})/E_{p_\beta} \big )\Bigg ]}{ \Big (i \Omega_n + \alpha E_{p_\beta}\Big )},
\end{eqnarray}
where $\alpha= \pm 1$ represents lower and upper bands respectively,  $\vec{p}_{\eta \beta}=p_x \hat{e}_1 +\eta p_y \hat{e}_2 + \Delta_{\eta,\beta} \hat{e}_3$,  
and $E_{p_{\eta\beta}}= |\epsilon_{p_{\eta \beta}}|$.
Following the usual procedure for  frequency summation, followed by the analytical continuation $i\omega\rightarrow \omega+i0^+$, the  polarization function of the $\eta$ valley  acquires the form,
\begin{eqnarray}
\Pi^{\eta}_{ij}(q,\omega) = -\frac{1}{4}\int \frac{d^2p}{(2\pi)^2}\sum_{\substack{\alpha,\alpha'=\pm1\\\beta,\beta'=\pm1}}\Big[F^{\beta,\beta'}_{i,j}\cdot S_{p,p+q}^{\alpha,\alpha',\beta,\beta'} \Big]  \nonumber\\ \times \frac{n_F(-\alpha E_{p_{\eta\beta}})-n_F(-\alpha'E_{(p+q)_{\eta\beta'}})}{ \Big (\alpha E_{p_{\eta\beta}}-\alpha' E_{(p+q)_{\eta\beta'}}-\omega - i 0^+\Big )},\label{eq:PF1}
\end{eqnarray}
where the prefactors are, $F^{\beta,\beta'}_{i,j}=\Big[ \delta_{ij}(1-\beta \beta')+i \epsilon_{izj}( \beta- \beta')+2\beta \beta' \delta_{iz}\delta_{jz} \Big]$ with $(i,j)\in (x,y,z)$, $F^{\beta,\beta'}_{0,j}=F^{\beta,\beta'}_{j,0}=(\beta + \beta') \delta_{zj}$
and $F^{\beta,\beta'}_{0,0}=(1+\beta \beta')$. The form factor is givenby
 \begin{eqnarray}
 S_{p,p+q}^{\alpha,\alpha',\beta,\beta'}= \Bigg[1+ \alpha \alpha'\frac{\vec{p}_{\eta\beta} \cdot \big(\vec{p}+\vec{q} \big)_{\eta\beta'} }{E_{p_{\eta\beta}}E_{(p+q)_{\eta\beta'}}}\Bigg].\label{eq:formfactor}
 \end{eqnarray}
 The   full polarization function is given by the sum, $\Pi_{ij}(q,\omega) = \Pi^{+}_{ij}(q,\omega) + \Pi^{-}_{ij}(q,\omega)$.  We note 
that the off-diagonal components, $\Pi^\eta_{0z}(q,\omega)$ and   $\Pi^\eta_{xy}(q,\omega)$, 
are non-zero in individual valleys, however, they add up to zero upon including contributions from both the valleys. This could be understood in the following way:
for $\Pi^{\pm}_{xy}$ the allowed transitions are between $\Delta_1$ to  $\Delta_2$, and viceversa. Focussing only on, $\Delta_1\rightarrow\Delta_2$ (or $\Delta_2\rightarrow\Delta_1$) transition, all terms in the expression of $\Pi^{\pm}_{xy}$ remain the same except the $i(\beta'-\beta)$ term which has opposite signs for the two valleys, thus the cancellation. Similar arguments hold for the vanishing of  $\Pi^{\pm}_{0z}$ term after including contributions from both the valleys. On the other hand, the diagonal components obtain equal contributions from both the valleys.
\begin{figure}
\vspace{0.5cm}
\hspace{1cm}
\centering
\includegraphics[width=1.0\linewidth]{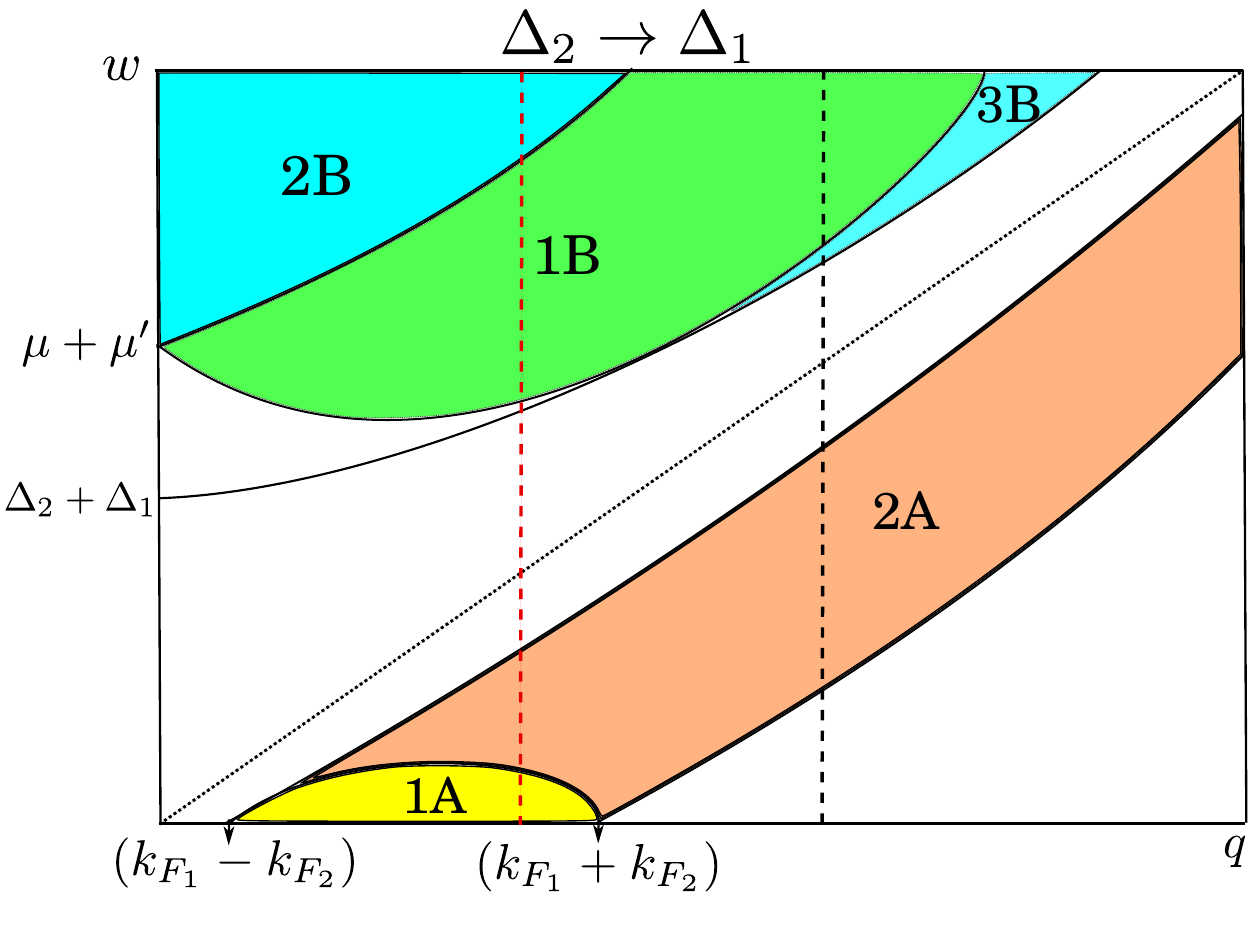}
\caption{(Color online) Shaded regions in the figure indicate non-zero contributions to the imaginary part of the polarization function  due to the $\Delta_2\rightarrow \Delta_1$ transition. $\text{1A}$ and $\text{2A}$ regions denote contributions from the transitions $u_{\Delta_2}$ to $u_{\Delta_1}$, whereas $\text{1B}$, $\text{2B}$ and $\text{3B}$ denote   contributions from $l_{\Delta_2}$ to $u_{\Delta_1}$. Here $\mu'=\sqrt{k_{F_1}^2+\Delta_2^2}$, $k_{F_1}=\sqrt{\mu^2-\Delta_1^2}$ and $k_{F_2}=\sqrt{\mu^2-\Delta_2^2}$.}
\label{Delta2-1}
\end{figure}

We will next focus our attention on the imaginary part of the polarization operator, in particular, those arising from  $\Pi_{xx}$ and $\Pi_{yy}$ (both of which yield identical result). The $\Pi_{00}$ result has already been discussed in the literature~\cite{pyat2009, andreas2012,nicol2014,peeters2014,anmol2016}, while the $\Pi_{zz}$ result follows trivially from those of the $\Pi_{00}$. The imaginary part of the polarization operator is non-zero in regions where the particle-hole excitations are allowed. For  $\Pi_{xx}$ and $\Pi_{yy}$, the contribution to their imaginary parts are obtained by  particle transition between bands with opposite spins ($\beta \beta'=-1$) and these could be due to transitions between  upper bands or from lower to upper band. For convenience, the bands are labeled as follows: upper and lower bands with band gap  $\Delta_{1/2} = |lE_z \mp \lambda_{SO}|$ as $u_{\Delta_{1/2}}$ and  $l_{\Delta_{1/2}}$, respectively. In the next two sub-sections, we will separately obtain contributions arising from $\Delta_{2/1} \rightarrow \Delta_{1/2}$ transitions, which when combined together give full contribution to $\Pi_{xx}$ and $\Pi_{yy}$. The calculations presented are for one of the valleys,  the other valley yields identical contribution.  Results are summarized below.

\begin{figure}
\centering
\includegraphics[width=1.0\linewidth]{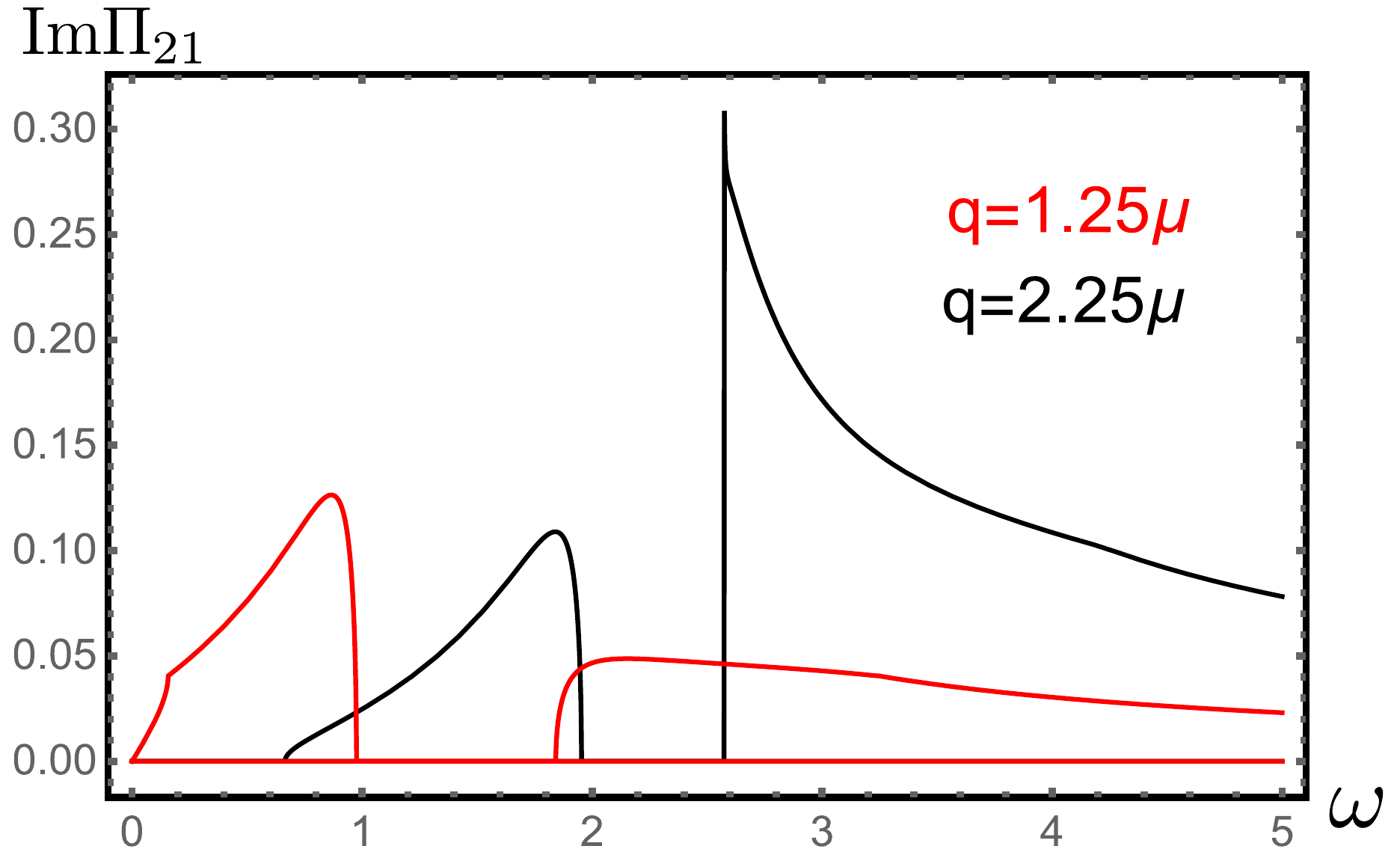}
\caption{(Color online)  Plotted are $\text{Im}\Pi_{21}$ vs $\omega$ for $k_{F_1}- k_{F_2}< q=1.25\mu<k_{F_1}+ k_{F_2} $ and  $k_{F_1}+ k_{F_2} < q=2.25\mu$. Here and in subsequent plots $\omega$ and $\Pi$ are in units of $\mu$.  }
\label{ImPi2-1}
\end{figure}

\subsection{ ($\Delta_2 \rightarrow \Delta_1$) Transition }
The transition from $u_{\Delta_2}$ to $u_{\Delta_1}$ is allowed for particles with energy $\epsilon$ in the range:
$ \text{max}[\mu-\omega,\Delta_2]< \epsilon< \mu$.  The angular integration of Eq.~\ref{eq:PF1} (with $\alpha=\alpha'=-1$) yields,
\begin{eqnarray}
\text{Im}\Pi_{21}^{uu}(q,\omega) =- \text{Re}\Bigg[\frac{1}{\sqrt{q^2-\omega^2}}\int_{L_x}^{U_x} \frac{dx}{8\pi}\frac{(x-\omega_1)^2-\gamma_0}{\sqrt{x^2-\xi_{21}^2}} \Bigg],\nonumber
\end{eqnarray} 
where  $\gamma_0=q^2+\Delta_d^2$, $\omega_1=\omega(\gamma_{21}-1)$, $\xi_{21}=\sqrt{q^2 \gamma_{21}^2+4q^2\Delta_2^2/(q^2-\omega^2)}$, $\gamma_{21}=1-\Delta_s\Delta_d/(q^2-\omega^2)$, along with the redefine parameter $\Delta_s=\Delta_2+\Delta_1$ and $\Delta_d=\Delta_2-\Delta_1$. Performing the integration by taking  the limits of integration to be   $U_x =2\mu +\omega \gamma_{21}$ and $L_x=2\text{max}[\mu-\omega,\Delta_2] +\omega \gamma_{21} $,  we obtain
\begin{widetext}
\[
\text{Im}\Pi_{21}^{uu}(q,\omega) =-\frac{1}{4\pi} \frac{1}{\sqrt{q^2-\omega^2}}\,\,\,\times \left \{
\begin{tabular}{ccc}
$G_{21}^{uu}\big(2\mu+\omega\gamma_{21}\big)-G_{21}^{uu}\big(2~\text{max}[\mu-\omega,\Delta_2]+\omega\gamma_{21}\big) \hspace{1cm}:$1A\\
$G_{21}^{uu}\big(2\mu+\omega\gamma_{21}\big)-G_{21}^{uu}\big(\xi_{21}\big) \hspace{4.3cm}:$2A
\end{tabular}
\right \},
\]
where
\begin{eqnarray}
G^{uu}_{21}(x)=\frac{1}{4}\Bigg\{\Big[-2q^2-2\Delta_d^2+\xi_{21}^2+2(\omega\gamma_{21}-\omega)^2\Big]
\log\big(\sqrt{x^2-\xi_{21}^2}+x\big)+\Big[x-4(\omega\gamma_{21}-\omega)\Big]\sqrt{x^2-\xi_{21}^2}\Bigg\}.
\end{eqnarray}
\end{widetext}
The  regions  in the $(q,\omega)$ plane  where $\text{Im}\Pi_{21}^{uu}(q,\omega)$  is non-zero are [see Fig.~(\ref{Delta2-1})]:
\begin{eqnarray}
&&1A :  \omega< \mu-\mathcal{F}(k_{F_1},\Delta_2) \nonumber\\
&&2A :  \pm\mu\mp \mathcal{F}(k_{F_{1(2)}},\Delta_{2(1)})<\omega<-\mu+\mathcal{F}(-k_{F_2},\Delta_1),\nonumber
\end{eqnarray}
where $\mathcal{F}(x,y)=\sqrt{(q-x)^2+y^2}$. 
The allowed regions for particle-hole (p-h) excitation in the $(q,\omega)$ plane can be obtained via kinematic consideration (see $\omega<q$ region in Fig.~\ref{Delta2-1}).
For example, in the scenario being discussed, the minimum momentum required for p-h generation is $k_{F_1}- k_{F_2}$, this involves the collinear transition of a particle from the  Fermi level of   $u_{\Delta_2}$ to the Fermi level of  $u_{\Delta_1}$ without a change in energy. Indeed, the particle's energy need not change for the transition from the 
Fermi-level of one band to the Fermi-level of the other band, thus the maximum momentum change for such a process  is $k_{F_2}+ k_{F_1}$.
For a given momentum $q > k_{F_1}- k_{F_2}$, the   energy upper bound for a transition from $u_{\Delta_2}$ to  $u_{\Delta_1}$ is $\omega_{\text{max}}= \sqrt{(k_{F_2}+q)^2 + \Delta_1^2 }-\mu$.  The process involves  a particle getting excited   from the Fermi level of  $u_{\Delta_2}$ to an unoccupied level of $u_{\Delta_1}$ with the final direction being the same as the initial one. On the other hand the lower boundary (for $q > k_{F_1}+ k_{F_2}$) is set by transition involving back-scattering of particle from the Fermi-level of $u_{\Delta_2}$ to $u_{\Delta_1}$ (with momentum change $ q-k_{F_2}$) which requires $\omega_{\text{min}}= \sqrt{(k_{F_2}-q)^2 + \Delta_1^2 }-\mu$.

A lower, $l_{\Delta_2}$, to upper band  $u_{\Delta_1}$ transition requires the particle to have energy $\epsilon$ in the range: $  \mu-\omega< \epsilon  < -\Delta_2$. Performing the angular integration of Eq.~\ref{eq:PF1} yields
\begin{eqnarray}
\text{Im}\Pi^{lu}_{21}(q,w) =- \text{Re}\Bigg[\frac{1}{\sqrt{\omega^2-q^2}}\int_{L_x}^{U_x} \frac{dx}{8\pi}\frac{\gamma_0-(x+w_1)^2}{\sqrt{\xi_{21}^2-x^2}} \Bigg],\nonumber
\end{eqnarray} 
where the limits of integration are  $U_x =2(\omega-\mu)-\omega \gamma_{21}$
and $L_x=2\Delta_2-\omega \gamma_{21} $. Integrating the above equation we obtain the following result:
\begin{widetext}
\[
\text{Im}\Pi^{lu}_{21}(q,\omega) =-\frac{1}{4\pi} \frac{1}{\sqrt{\omega^2-q^2}}\,\,\,\times \left \{
\begin{tabular}{ccc}
$G_{21}^{lu}\big(2(\omega-\mu)-\omega\gamma_{21}\big)-G_{21}^{lu}\big(-\xi_{21}\big) \hspace{0.8cm}:$1B \nonumber\\
$G_{21}^{lu}\big(\xi_{21}\big)-G_{21}^{lu}\big(-\xi_{21}\big)\hspace{2.8cm} :$2B \nonumber\\
$G_{21}^{lu}\big(\xi_{21}\big)-G_{21}^{lu}\big(-\xi_{21}\big)\hspace{2.8cm} :$3B \nonumber
\end{tabular}
\right \},
\]
where,
\begin{eqnarray}
G^{lu}_{21}(x)=\frac{1}{4}\Bigg\{\Big[2q^2+2\Delta_d^2-\xi_{21}^2-2(\omega\gamma_{21}-\omega)^2\Big]
\tan^{-1}\Big(\frac{x}{\sqrt{\xi_{21}^2-x^2}}\Big)  +\Big[x-4(\omega\gamma_{21}-\omega)\Big]\sqrt{\xi_{21}^2-x^2}\Bigg\}.
\end{eqnarray}
	
\end{widetext}	
The non-zero  regions in the $(q,\omega)$ plane  are described by the following equations
\begin{eqnarray}
&&1B :  \mu+\mathcal{F}(k_{F_1},\Delta_2)<\omega<\mu+\mathcal{F}(-k_{F_1},\Delta_2) \hspace{5.25cm}\nonumber\\
&&2B :  \omega>\mu+\mathcal{F}(-k_{F_1},\Delta_2)  \hspace{4.95cm}  \nonumber\\
&&3B :  \sqrt{q^2+\Delta_s^2}  <\omega< \mu+\mathcal{F}(k_{F_1},\Delta_2) \hspace{3.75cm}\nonumber
\end{eqnarray}

Unlike the transitions involving only the upper bands, $q=0$ particle-hole transitions are now allowed 
for all frequencies $\omega>\mu + \sqrt{k_{F_1}^2 + \Delta_2^2}$ (see $\omega>q$ region in Fig.~\ref{Delta2-1}). As $q$ is increased, the threshold  frequency  given by   $\omega=\mu + \sqrt{(k_{F_1}-q)^2 + \Delta_2^2}$ exhibits a downturn, these are realized by processes involving particle with momentum $p<k_{F_1}$  moving to the upper Fermi level while maintaining its initial direction. For the above process, the minimum allowed frequency $\omega=\mu + \Delta_2$ is reached for $q=k_{F_1}$, where the transitioning particle had originally momentum $p=0$. Increasing $q$ further,
the threshold frequency exhibits an upturn. The process  now involves particle from $l_{\Delta_2}$ moving to the upper Fermi level by changing  its initial direction. A further increase in  $q$ changes the threshold frequency to $\omega =\sqrt{q^2 + \Delta_s^2}$ and is obtained by minimizing $\sqrt{(p-q)^2 + \Delta_2^2}+\sqrt{p^2 + \Delta_1^2}$ with respect to $p$. 

Combining $\text{Im}\Pi_{21}^{uu}$ and $\text{Im}\Pi_{21}^{lu}$ yields the  contribution to the imaginary part of the polarization operator from the $2\rightarrow 1$ processes represented as $\text{Im}\Pi_{21}$.   In Fig.~\ref{ImPi2-1} we have plotted $\text{Im}\Pi_{21}$ as a function of $\omega$ for two  values of $q$. The frequencies for which $\text{Im}\Pi_{21}$ vanishes represent regions for which single p-h excitations are forbidden.
For $l_{\Delta_2}\rightarrow u_{\Delta_1}$ transition (right most curves of Fig.~\ref{ImPi2-1}), the threshold behavior  	exhibits contrasting features depending on whether 
 $q$ is lesser or greater than  
 $(\Delta_s +\sqrt{\mu^2-\Delta{_2}^2})/\Delta_1$ 
 (the value at which $\omega = \sqrt{q^2 +\Delta_s^2} $ and $\omega = \mu + \sqrt{(q-k_{F_1})^2+ \Delta_2^2}$ curves intersect).   For $q$  values greater than  $q^*=(\Delta_s+\sqrt{\mu^2-\Delta_2^2})/\Delta_1$   the threshold behavior exhibits a step jump (shown  by the black curve) to a finite value given by $q^2\Delta_1\Delta_2/\Delta_s^3$, whereas for  lesser values of $q$ it vanishes with the derivative acquiring a square-root singularity  at $\omega = \mu + \sqrt{(q-k_{F_1})^2+ \Delta_2^2}$  (shown  by the  red curve). 
On the other hand, for $u_{\Delta_2}$ to $  u_{\Delta_1}$ transition,   the threshold behavior at the upper edge of region $2A$  vanishes, while  the derivative diverges again with square-root singularity. Moreover,  inside the allowed regions the plot exhibits a weak kink 
at various boundaries.

\begin{figure}
\centering
\includegraphics[width=0.9\linewidth]{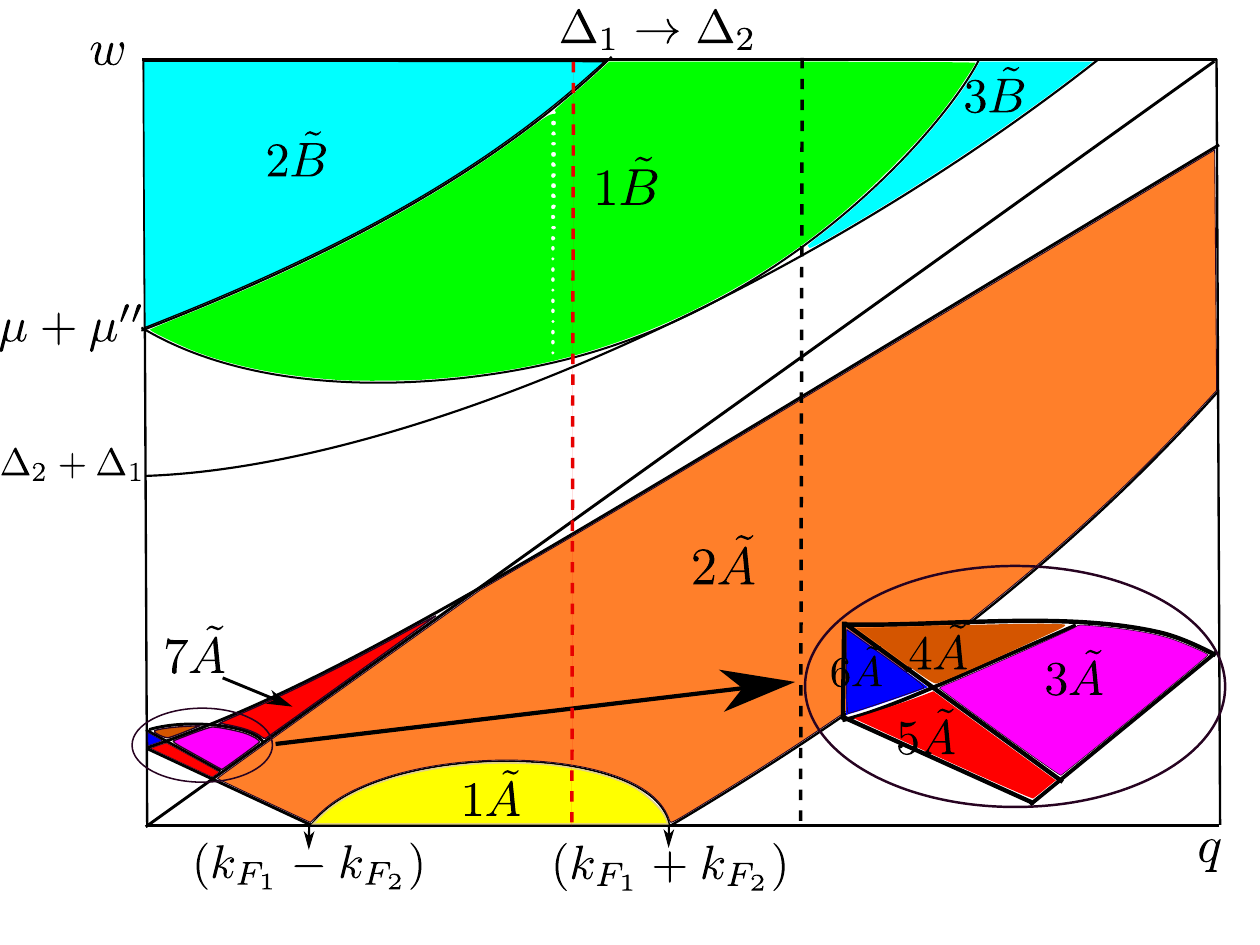}
\caption{(Color online) Regions in the $(q,\omega)$ plane where $\Delta_1\rightarrow \Delta_2$ transitions contribute to the imaginary part of the polarization function. $\tilde{\text{A}}$ and $\tilde{\text{B}}$ regions denote contributions from  $u_{\Delta_1}$ to $u_{\Delta_2}$ and   $l_{\Delta_1}$ to $u_{\Delta_2}$ transitions, respectively. Here $\mu''=\sqrt{k_{F_2}^2+\Delta_1^2}$.}  
\label{Delta1-2}
\end{figure}
\subsection{($\Delta_1 \rightarrow \Delta_2$) Transition }
Similar to the earlier discussed upper band transitions, the  transition from $u_{\Delta_1}$ to $u_{\Delta_2}$ are allowed for particles with energy $\epsilon$ in the range: 
$ \text{max}[\mu-\omega,\Delta_1]< \epsilon< \mu$.  The major difference is that now the particle-hole transitions are allowed even for $\omega > q$ regions, albeit the phase-space is much smaller than the phase space for the dominant $\omega<q$ regions [see the lower part of the $(q,\omega)$ plane in Fig.~\ref{Delta1-2}].

The maximum allowed frequency  for such a transition is given by $\omega_{\text{max}}=\text{max}[\mu-\sqrt{(k_{F2}-q)^2 +\Delta^2_1},\sqrt{(k_{F1}+q)^2 +\Delta^2_2}-\mu]$. The first term in the square brackett is the energy $\mu-\sqrt{(k_{F2}-q)^2 +\Delta^2_1}$ required for a  colinear  transition of a particle from $u_{\Delta_1}$ to the Fermi level of $u_{\Delta_2}$. These transitions serve as the upper bound for frequency at small momentum transfer.  The second frequency term $\sqrt{(k_{F1}+q)^2 +\Delta^2_2}-\mu $ is due to the collinear transition of a particle to $u_{\Delta_2}$ originating  from  the Fermi-level of $u_{\Delta_1}$. 
The lower bound of frequency for the $u_{\Delta_1}$ to $u_{\Delta_2}$ transition include 
$ \sqrt{(k_{F1}-q)^2 +\Delta^2_2}-\mu$ (collinear transition from the Fermi level of the first band to the second band with the reduced momentum of the final particle) for momentum exchanges which lie between $0<q< k_{F1}-k_{F2}$. In the range $ k_{F1}-k_{F2}<q< k_{F1}+k_{F2}$ the transition can take place without change in the  energy of the particle. While in the range 
$k_{F1}+k_{F2} < q$ the minimum energy required is $ \sqrt{(k_{F1}-q)^2 +\Delta^2_2}-\mu$, which involves a transition from the Fermi level of the first band to a higher energy level of the second band with the final momentum reversing its direction.

\begin{figure}
\vspace{0.5cm}
\centering
\includegraphics[width=1.0\linewidth]{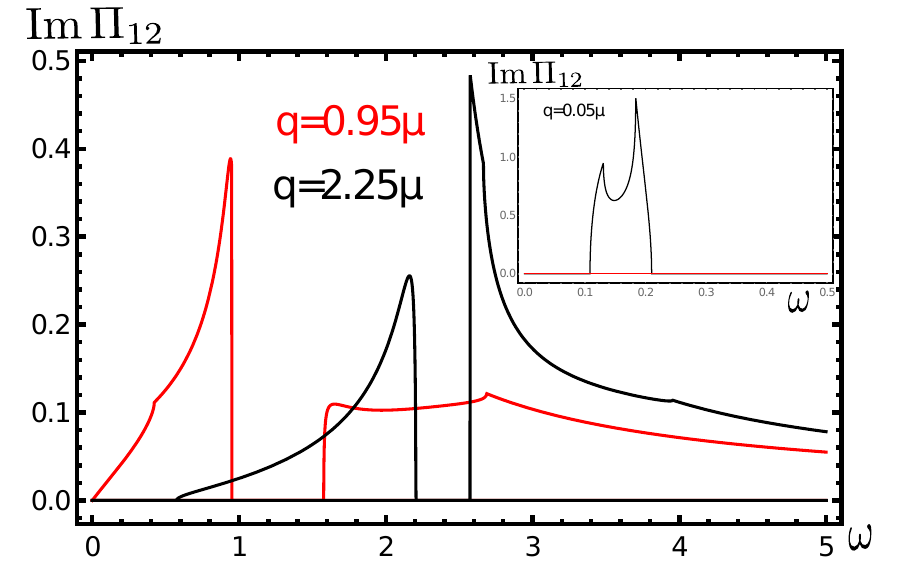}
\caption {(Color online)  Plotted are $\text{Im}\Pi_{12}$ vs $\omega$ for $q=0.95\mu, \text{and}~ 2.25\mu $, where  $k_{F_1}- k_{F_2}< 0.95\mu<k_{F_1}+ k_{F_2} $ and  $k_{F_1}+ k_{F_2} < q=2.25\mu$. The features are very similar to those shown in  Fig.~\ref{ImPi2-1}, except here the discontinuities in the slopes are more pronounced.  In the inset we plot for $q=0.05\mu$, where $0.05\mu< k_{F_1}- k_{F_2}$ , this additional feature is unique to $\Delta_1 \rightarrow \Delta_2$ transitions.  }
\label{ImPi1-2}
\end{figure}
The  contribution to the imaginary part of the  polarization function  are as follows:
\begin{widetext}
\[
\text{Im}\Pi^{uu}_{12}(q,\omega) =-\frac{1}{4\pi} \frac{1}{\sqrt{|q^2-\omega^2|}}\,\,\,\times \left \{
\begin{tabular}{ccc}
$G^{uu}_{12}\big(2\mu+\omega\gamma_{12}\big)-G^{uu}_{12}\big(2\text{max}[\mu-\omega,\Delta_1]+\omega\gamma_{12}\big) \hspace{1.4cm}:1\tilde{A}$\\
$G^{uu}_{12}\big(2\mu+\omega\gamma_{12}\big)-G^{uu}_{12}\big(\xi_{12}\big) \hspace{4.6cm}:2\tilde{A}$\\
$\bar{G}^{uu}_{12}\big(2\mu+\omega\gamma_{12}\big)-\bar{G}^{uu}_{12}\big(2\text{max}[\mu-\omega,\Delta_1]+\omega\gamma_{12}\big) \hspace{1.4cm}:3\tilde{A}$\\
$\bar{G}^{uu}_{12}\big(\xi_{12}\big)-\bar{G}^{uu}_{12}\big(2\text{max}[\mu-\omega,\Delta_1]+\omega\gamma_{12}\big) \hspace{2.5cm}:4\tilde{A}$\\
$\bar{G}^{uu}_{12}\big(2\mu+\omega\gamma_{12}\big)-\bar{G}^{uu}_{12}\big(-\xi_{12}\big)\hspace{4.2cm} :5\tilde{A}$\\
$\bar{G}^{uu}_{12}\big(\xi_{12}\big)-\bar{G}^{uu}_{12}\big(-\xi_{12}\big)\hspace{5.3cm} :6\tilde{A}$\\
$\bar{G}^{uu}_{12}\big(2\mu+\omega\gamma_{12}\big)-\bar{G}^{uu}_{12}\big(-\xi_{12}\big)\hspace{4.2cm} :7\tilde{A} $
\end{tabular}
\right \},
\]
where $\gamma_{12}=1+\Delta_s\Delta_d/(q^2-\omega^2)$, $\xi_{12}=\sqrt{q^2 \gamma_{12}^2+4q^2\Delta_1^2/(q^2-\omega^2)}$ and 
\begin{eqnarray}
G^{uu}_{12}(x)&=&\frac{1}{4}\Bigg\{\Big[-2q^2-2\Delta_d^2+\xi_{12}^2+2(\omega\gamma_{12}-\omega)^2\Big]	\log\big(\sqrt{x^2-\xi_{12}^2}+x\big) +\Big[x-4(\omega\gamma_{12}-\omega)\Big]\sqrt{x^2-\xi_{12}^2}\Bigg\},\\
\bar{G}^{uu}_{12}(x)&=&\frac{1}{4}\Bigg\{\Big[-2q^2-2\Delta_d^2+\xi_{12}^2+2(\omega\gamma_{12}-\omega)^2\Big]   \tan^{-1}\big(\frac{x}{\sqrt{\xi_{12}^2-x^2}}\big)     	
-\Big[x-4(\omega\gamma_{12}-\omega)\Big]\sqrt{\xi_{12}^2-x^2}\Bigg\}.
\end{eqnarray}
The different allowed regions in the $(q,\omega)$ plane for the    $u_{\Delta_1}$ to $u_{\Delta_2}$ transition (Fig.~\ref{Delta1-2}) are as follows, 
\begin{eqnarray}
&&1\tilde{A} :  \omega<\mu-\mathcal{F}(k_{F_2},\Delta_1), \nonumber\\
&&2\tilde{A} :  \pm\mu\mp \mathcal{F}(k_{F_{2(1)}},\Delta_{1(2)})<\omega<-\mu+\mathcal{F}(-k_{F_1},\Delta_2),\hspace{4.95cm}\nonumber\\
&&3\tilde{A} :  \omega>q;\,\, \& \,\,\omega<\mu-\mathcal{F}(k_{F_2},\Delta_1);\,\, \& \,\,\omega>\mu-\mathcal{F}(-k_{F_2},\Delta_1);\,\, \& \,\,\omega<-\mu+\mathcal{F}(-k_{F_1},\Delta_2),\nonumber\\
&&4\tilde{A} :  \omega>q;\,\, \& \,\,\omega<\mu-\mathcal{F}(k_{F_2},\Delta_1);\,\, \& \,\,\omega>\mu-\mathcal{F}(-k_{F_2}),\Delta_1);\,\, \& \,\,\omega>-\mu+\mathcal{F}(-k_{F_1},\Delta_2),\nonumber\\
&&5\tilde{A} :  \omega>q;\,\, \& \,\,\omega<-\mu+\mathcal{F}(-k_{F_1},\Delta_2);\,\, \& \,\,\omega>-\mu+\mathcal{F}(k_{F_1},\Delta_2);\,\, \& \,\,\omega<\mu-\mathcal{F}(-k_{F_2},\Delta_1),\nonumber\\
&&6\tilde{A} :  \omega<\mu-\mathcal{F}(-k_{F_2},\Delta_1);\,\, \& \,\,\omega>-\mu+\mathcal{F}(-k_{F_1},\Delta_2),\nonumber\\
&&7\tilde{A} :  \omega>q;\,\, \& \,\,\omega>\mu-\mathcal{F}(k_{F_2},\Delta_1);\,\, \& \,\,w<-\mu+\mathcal{F}(-k_{F_1},\Delta_2),\nonumber
\end{eqnarray}
\end{widetext}

A lower band $l_{\Delta_1}$ to upper band  $u_{\Delta_2}$ transition  requires the particle to have energy $\epsilon$ in the range: $  \mu-\omega< \epsilon  < -\Delta_1$. The derivation of the threshold frequencies are very similar as 
for the case of $l_{\Delta_2}$ to $u_{\Delta_1}$ transition and are obtained by simply exchanging the indices $1\rightleftharpoons 2$.  The threshold frequency for small $q$ has the form  $\omega=\mu + \sqrt{(k_{F_2}-q)^2 + \Delta_1^2}$ which changes to  $\omega =\sqrt{q^2 + \Delta_s^2}$ at the point of intersection of the two curves.  The contribution to the imaginary part of the polarization function are obtained to be:

\begin{widetext}
\[
\text{Im}\Pi^{lu}_{12}(q,\omega) =-\frac{1}{4\pi} \frac{1}{\sqrt{\omega^2-q^2}}\,\,\,\times \left \{
\begin{tabular}{ccc}
$G_{12}^{lu}\big(2(\omega-\mu)-\omega\gamma_{12}\big)-G_{12}^{lu}\big(-\xi_{12}\big) \hspace{1.35cm}:1\tilde{B}$ \nonumber\\
$G_{12}^{lu}\big(\xi_{12}\big)-G_{12}^{lu}\big(-\xi_{12}\big)\hspace{3.4cm} :2\tilde{B}$ \nonumber\\
$G_{12}^{lu}\big(\xi_{12}\big)-G_{12}^{lu}\big(-\xi_{12}\big)\hspace{3.4cm} :3\tilde{B}$ \nonumber
\end{tabular}
\right \},
\]
where,
\begin{eqnarray}
G^{lu}_{12}(x)=\frac{1}{4}\Bigg\{\Big[2q^2+2\Delta_d^2-\xi_{12}^2-2\big(\omega\gamma_{12}-\omega\big)^2\Big]
\tan^{-1}\big(\frac{x}{\sqrt{\xi_{12}^2-x^2}}\big)+\Big[x-4\big(\omega\gamma_{12}-\omega\big)\Big]\sqrt{\xi_{12}^2-x^2}\Bigg\}.
\end{eqnarray}
\end{widetext}
The non-zero  regions  in the $(q,\omega)$ plane (Fig.~\ref{Delta1-2}) are,
\begin{eqnarray}
&&1\tilde{B} :  \mu+\mathcal{F}(k_{F_2},\Delta_1)<\omega<\mu+\mathcal{F}(-k_{F_2},\Delta_1) \hspace{5.25cm}\nonumber\\
&&2\tilde{B} :  \omega>\mu+\mathcal{F}(-k_{F_2},\Delta_1)  \hspace{5.0cm}  \nonumber\\
&&3\tilde{B} :  \sqrt{q^2+(\Delta_2+\Delta_1)^2}  <\omega< \mu+\mathcal{F}(k_{F_2},\Delta_1).\hspace{3.95cm}\nonumber
\end{eqnarray}
Fig.~\ref{ImPi1-2} shows $\text{Im}\Pi_{12} =\text{Im}\Pi_{12}^{uu}+\text{Im}\Pi_{12}^{lu} $ plotted as a function of  $\omega$ for three different values of $q$. 
The behavior for  $l_{\Delta_1}\rightarrow u_{\Delta_2}$  (right most curves of Fig.~\ref{ImPi1-2}) transition is similar to those  considered in  Fig.~\ref{ImPi2-1}.
In this case, the main change is in the position of 
$q$ value given by $q^*=(\Delta_s+\sqrt{\mu^2-\Delta_1^2})/\Delta_2$ which separates the two threshold behaviors. As before, for $q$ values greater than it, the threshold behavior exhibits a step jump to  the same finite value $q^2\Delta_1\Delta_2/\Delta_s^3$ (shown  by the black curve), whereas for lesser $q$ values the derivative at the threshold diverges (shown  by the red curve).
Also, for  $u_{\Delta_1}$ to $  u_{\Delta_2}$ transition,   the threshold behavior at the upper edge   vanishes everywhere, while  the derivative diverges with square-root singularity. For the additional region shown in the inset, at small $\omega$ and $q< k_{F_1} -k_{F_2}$, the threshold behavior at both the edges  exhibits square-root divergence of the derivatives.  It turns out that in  this region the real part of the polarization operator exhibits singular  features, details of which are provided in sec.~\ref{sec: Real_q_0}. Finally to conclude this section, $\text{Im}\Pi_{xx/yy}$  is given by  
$\text{Im}\Pi_{xx/yy}= \text{Im}\Pi_{21}+\text{Im}\Pi_{12}$.
It is worth mentioning that in the absence of electric-field, $\Delta_1
=\Delta_2$, therefore $\text{Im}\Pi_{12}$  and  $\text{Im}\Pi_{21}$
will be identical.



\subsection{ ($\Delta_{1(2)} \rightarrow \Delta_{1(2)}$ ) Transition }
For completeness we will enumerate the known result corresponding to the case of intra and inter-band transitions within  the same gap, $i.e.,$ $\Delta_{i} \rightarrow \Delta_{i}$ where $i\in (1,2)$~\cite{pyat2009}. 
 These  give contributions to only  $\text{Im} \Pi_{zz}$ and as before they arise due to $u_{\Delta_i}\rightarrow u_{\Delta_i}$ and $l_{\Delta_i}\rightarrow u_{\Delta_i}$
transitions.  The contribution to the  imaginary part of the polarization function  from the $u_{\Delta_i}\rightarrow u_{\Delta_i}$ transitions are as follows,
\begin{figure}
\centering
\includegraphics[width=0.9\linewidth]{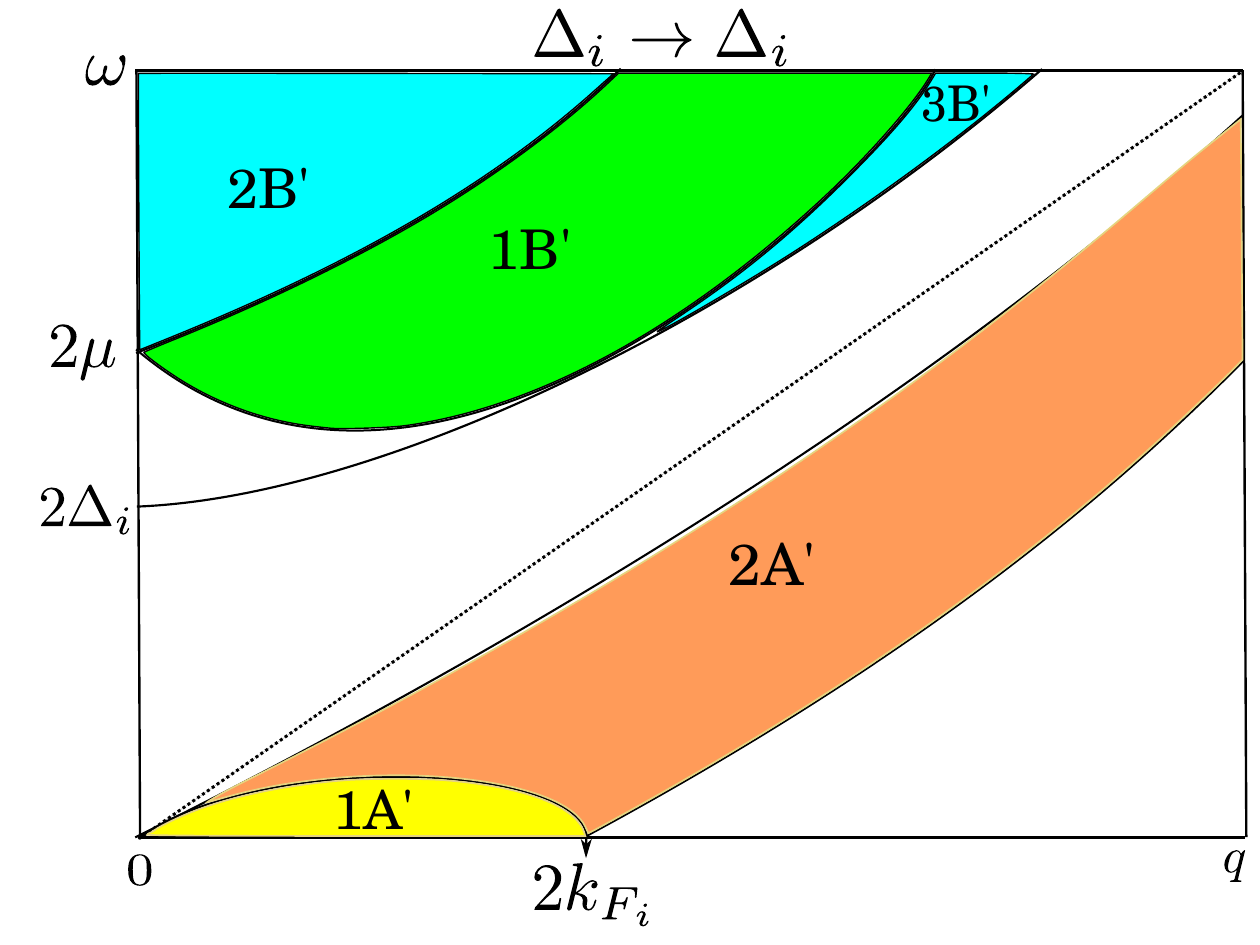}
\caption{(Color online) The $\text{A}'$ regions denote contributions from $u_{\Delta_i}\rightarrow u_{\Delta_i}$, whereas $\text{B}'$  denote those  from $l_{\Delta_i}\rightarrow u_{\Delta_i}$. Note $k_{F_i}=\sqrt{\mu^2-\Delta_i^2}$.}
\label{Fig4}
\end{figure}
\begin{widetext}	
\[
\text{Im}\Pi^{uu}_{ii}(q,\omega) =-\frac{1}{4\pi} \frac{1}{\sqrt{q^2-\omega^2}}\,\,\,\times \left \{
\begin{tabular}{ccc}
$G^{uu}\big(2\mu+\omega\big)-G^{uu}\big(2\text{max}[\mu-\omega,\Delta_i]+\omega\big) \hspace{1.0cm}:1A'$ \\
$G^{uu}\big(2\mu+\omega\big)-G^{uu}\big(\xi\big) \hspace{0.5cm} \hspace{3.35cm}:2A'$
\end{tabular}
\right \},
\]
\begin{eqnarray}
{\text{where}}\,\,\xi=\sqrt{q^2 +4q^2\Delta_i^2/(q^2-\omega^2)},\quad{\text{and}}\quad G^{uu}(x)=\frac{1}{4}\Bigg\{\Big[\xi^2-2q^2\Big]\log\big(\sqrt{x^2-\xi^2}+x\big) +x\sqrt{x^2-\xi^2}\Bigg\}.
\end{eqnarray}	
\end{widetext}
The allowed regions for the transitions are (see Fig.~\ref{Fig4})
\begin{eqnarray}
&&1A' :  \omega<\mu-\mathcal{F}(k_{F_i},\Delta_i) \nonumber\\
&&2A' :  \pm\mu\mp \mathcal{F}(k_{F_i},\Delta_i)<\omega<-\mu+\mathcal{F}(-k_{F_i},\Delta_i).\hspace{4.95cm}\nonumber
\end{eqnarray}
Unlike the earlier two cases, the  transitions within the same band allows the creation of particle-hole pairs having $\omega=0$  and infintesimally small momentum $q$.

The  contribution  from  $l_{\Delta_i}\rightarrow u_{\Delta_i}$ transitions are,
\begin{widetext}
\[
\text{Im}\Pi^{lu}_{ii}(q,\omega)=-\frac{1}{4\pi} \frac{1}{\sqrt{\omega^2-q^2}}\,\,\,\times \left \{
\begin{tabular}{ccc}
$G^{lu}\big(\omega-2 \mu\big)-G^{lu}\big(-\xi\big) \hspace{1.8cm}:1B'$ \\
$G^{lu}\big(\xi\big)-G^{lu}\big(-\xi\big) \hspace{2.75cm}:2B'$ \\
$G^{lu}\big(\xi\big)-G^{lu}\big(-\xi\big) \hspace{2.75cm}:3B'$\notag
\end{tabular}
\right \},
\]
where
\begin{eqnarray}
G^{lu}(x)=\frac{1}{4}\left[\big(2q^2-\xi^2\big)  \tan^{-1}\bigg(\frac{x}{\sqrt{\xi^2-x^2}}\bigg) +x\sqrt{x^2-\xi^2}\right]
\end{eqnarray}
and the allowed regions in the  $(q,\omega)$ plane are 
(see Fig.~\ref{Fig4}) 
\begin{eqnarray}
&&1B' :  \mu+\mathcal{F}(k_{F_i},\Delta_i)<\omega<\mu+\mathcal{F}(-k_{F_i},\Delta_i)\nonumber\\
&&2B' :  \omega>\mu+\mathcal{F}(-k_{F_i},\Delta_i)  \nonumber\\
&&3B' :  \omega>(2k_{F_i});\,\, \& \,\,\sqrt{q^2+(2\Delta_i)^2} <\omega< \mu+\mathcal{F}(k_{F_i},\Delta_i).\notag
\end{eqnarray}
\end{widetext}
We note that the qualitative behavior of this region is similar to the earlier two studied cases. As an additional remark, we would like to point out that in the scenario of vanishing electric field, the $zz$ component obtains identical contribution to $xx/yy$ components.\\


\section{ Real part of spin-susceptibility}\label{sec: Real_q_0}
The real part of spin-susceptibility is evaluated from  Eq.~\ref{eq:PF1}, where some of the parts have been calculated with the help of Kramers-Kronig technique and the rest via direct integration.  The Re$\chi_{xx}$ and   $\text{Re}\chi_{yy}$ are identical and obtain contributions from transitions involving $\Delta_1\rightarrow\Delta_2$ and viceversa, while $\Delta_i\rightarrow\Delta_i$ ($i=1,2$) transitions yield contributions to  $\text{Re}\chi_{zz}$. Details of the calculation are provided in appendix~\ref{appen}. In the following two subsections we will limit our discussion to the   case  of dynamic and static susceptibility. 
 
\subsection{Dynamic limit: $q=0$}
It is easy to show that for finite frequencies and $q=0$, $\text{Re}\chi^{0}_{zz}(q=0,\omega)$ vanishes identically due to the Fermi-distribution terms in (\ref{eq:PF1}) (for $\alpha=\alpha'$) and form factor (\ref{eq:formfactor}) (for $\alpha=-\alpha'$). 
In contrast,  $\text{Re}\chi^{0}_{xx}(0,\omega)$ and $\text{Re}\chi^{0}_{yy}(0,\omega)$ are in general  non-zero  and exhibit interesting behavior in regions where the corresponding imaginary part vanishes. 
In the following, we will take a closer look into the different contributions to the real 
part of the susceptibility. As before, we will discuss the susceptibility in terms of the polarization operator which differs by a sign.

The non-interacting real part of the polarization operator (the $xx$ and $yy$ components) is split  in to three parts labelled as  $\text{Re}\Pi_a$, $\text{Re}\Pi_b$  and $\text{Re}\Pi_c$ (details of the decomposition and their derivation are given in the appendix~\ref{appen_C}).  The first part, $\text{Re}\Pi_a$, is independent of $\mu$ and takes on the value,
\begin{eqnarray}
{\text{Re}}\,\Pi_a(\omega)
=-\frac{\Delta_d^2}{4\pi\omega}\Bigg\{\log\bigg[\frac{\Delta_s+\omega}{|\Delta_s-\omega|}\bigg]\Bigg(1-\frac{\Delta_s^2}{\omega^2}\Bigg)+\frac{2\Delta_s}{\omega}\Bigg\}\notag
\end{eqnarray}
where  $\Delta_d=\Delta_2-\Delta_1$ and $\Delta_s=\Delta_1+\Delta_2$. The second term, $\text{Re}\Pi_b$  is non-zero for $\mu > \Delta_1$ and obtains contribution from the integrals containing $n_F(\sqrt{p^2+ \Delta_1^2}-\mu)$ and $n_F(\sqrt{(p+q)^2+ \Delta_1^2}-\mu)$ terms and  is given by,
\begin{widetext}
\begin{eqnarray}
{\text{Re}}\Pi_{b}(\omega)
=-\frac{1}{4\pi} \Bigg\{ \frac{\Delta_d^2 \left(\Delta_s^2-\omega ^2\right)}{2\omega^3}\left( \log \left[\frac{\left(-\Delta_d\Delta_s-2\mu\omega+\omega ^2\right)   \left(-\Delta_d\Delta_s+2\omega\Delta_1+\omega ^2\right)}{\left(-\Delta_d\Delta_s+2\mu\omega+\omega ^2\right)   \left(-\Delta_d\Delta_s-2\omega\Delta_1+\omega ^2\right)}\right] \right)-\frac{2\Delta_d\Delta_s(\mu-\Delta_1)}{\omega^2}\Bigg\}.\label{eq:RePi_b}
\end{eqnarray}
The third term denoted as $\text{Re}\Pi_c$ obtains contribution from the integrals containing $n_F(\sqrt{p^2+ \Delta_2^2}-\mu)$ and $n_F(\sqrt{(p+q)^2+ \Delta_2^2}-\mu)$ terms   and  exhibits $\log$ divergence. It  has the following form,
\begin{eqnarray}
{\text{Re}}\,\,\Pi_{c}(\omega)
=-\frac{1}{4\pi} \Bigg\{ \frac{\Delta_d^2 \left(\Delta_s^2-\omega ^2\right)}{2\omega^3}\left( \log \left[\frac{\left(\Delta_d\Delta_s-2\mu\omega+\omega ^2\right)\left(\Delta_d\Delta_s+2\omega\Delta_2+\omega ^2\right)}{\left(\Delta_d\Delta_s+2\mu\omega+\omega ^2\right)\left(\Delta_d\Delta_s-2\omega\Delta_2+\omega ^2\right)}\right] \right)+\frac{2\Delta_d\Delta_s(\mu-\Delta_2)}{\omega^2}\Bigg\}.\label{eq:RePi_c}
\end{eqnarray}
\end{widetext}
Combining all the contributions, 
${\text{Re}}\Pi^0(\omega)={\text{Re}}\Pi_a(\omega)+{\text{Re}}\Pi_b(\omega)+{\text{Re}}\Pi_c(\omega)$, we obtain the following  compact expression, 
\begin{eqnarray}
{\text{Re}}\Pi^0(\omega)
&=&\frac{\Delta_d^2 \left(\omega ^2- \Delta_s^2\right)}{8\pi\omega^3}L(w),\label{eq:RePi}
\end{eqnarray}
where
\begin{eqnarray}
L(w)
&=&\log \left[\frac{(\omega ^2-2\mu\omega)^2-\Delta_d^2\Delta_s^2}{(\omega ^2+2\mu\omega)^2-\Delta_d^2\Delta_s^2}\right]. \notag
\end{eqnarray}
Let us next consider the possibility of spin collective excitations occuring in  the  $xx$ and $yy$ channels when coupled with interactions. The ladder diagrams yield an equation for  spin collective excitations which typically has the form, $\text{Re}\Pi_{xx/yy}^{0}(\omega)=-1/u^*$ ($u^*$ is the screened interaction).  It is clear that  this equation is satisfied, if $\text{Re}\Pi_{xx/yy}^{0}(\omega)$ is negative and singular (for weak interactions). Moreover, the frequencies which satisfy the equation should be in the range where $\text{Im}\Pi_{xx/yy}^{0}(\omega)$  vanishes so that the absence of single particle excitations leave the collective excitations undamped.  There are two such regimes where $\text{Im}\Pi^{0}_{xx/yy}(0,\omega)=0$, these include $0<\omega<\sqrt{\mu^2 -\Delta_1^2 +\Delta_2^2} -\mu$ and $ \mu -\sqrt{\mu^2 +\Delta_1^2 -\Delta_2^2}<\omega< \mu +\sqrt{\mu^2  + \Delta_1^2 -\Delta_2^2}$, both the constraints   are determined  by the $\Delta_1\rightarrow \Delta_2$ transitions (see Fig.\ref{Delta1-2}). 

\begin{figure}
\centering
\includegraphics[width=1.0\linewidth]{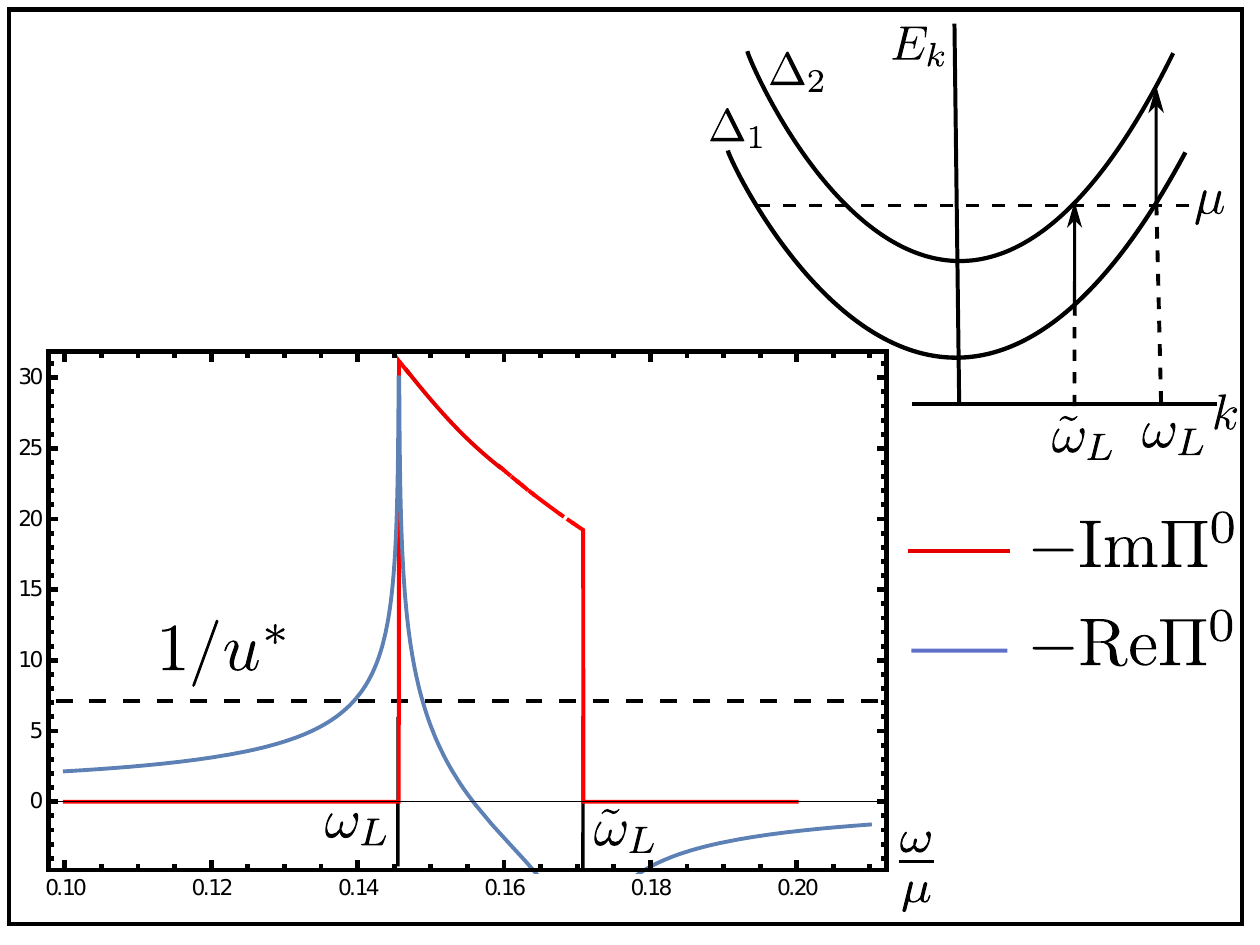}
\caption{(Color online) 
The dotted line denotes $1/u*$, while the blue curve represents $-\text{Re}\Pi^{0}(\omega)$. The collective excitation pole (near  the lower threshold) is given by the frequency at which they interesect. The red line corresponds to the imaginary part of the polarization function, where its boundaries are $\omega_L= \sqrt{\mu^2 -\Delta_1^2 +  \Delta_2^2 }-\mu$ and $\tilde{\omega}_L=\mu- \sqrt{\mu^2 +\Delta_1^2 - \Delta_2^2 }$}
\label{Pole_position}
\end{figure}

In the first interval, $0<\omega<\sqrt{\mu^2 -\Delta_1^2 +\Delta_2^2} -\mu$,  $\text{Re}\Pi_{xx/yy}^{0}(\omega)$ is negative  and has a logarithmic divergence right at the lower  threshold  of the single particle excitation, i.e., at $\omega_L = \sqrt{\mu^2 -\Delta_1^2 +\Delta_2^2} -\mu$.  
The log-divergence is manifested in $\text{Re}\Pi_{b}$, Eq.~\ref{eq:RePi_b},   because of the vanishing of the  first term in the denominator of the $\log$-term  at the frequency $\omega_L$.  The specific integral causing the divergence is
\begin{widetext}
\begin{eqnarray}
I\propto \int \frac{pdp}{8\pi}\left[\left(1+\frac{p^2+\Delta _1 \Delta_2}{\sqrt{p^2+\Delta _1^2} \, \sqrt{p^2+\Delta _2^2}}\right)\frac{\Theta
\left(\sqrt{\mu ^2-\Delta _1^2}-p\right)}{w+\sqrt{p^2+\Delta _1^2}-\sqrt{p^2+\Delta_2^2}}\right].
\end{eqnarray} 
\end{widetext}
One can deduce from the corresponding imaginary part  that the processes responsible for the contribution involve upper-band transitions from $\Delta_1\rightarrow\Delta_2$ as shown in Fig~\ref{Pole_position}.

Interestingly,  the real part of the polarization operator is also negative for frequencies $\omega$ close to and  less than $\omega_U= \mu + \sqrt{\mu^2 + \Delta_1^2 - \Delta_2^2}$ (the upper threshold   for the single particle excitation) and is logarithmically divergent right at $\omega=\omega_U$. The log-divergence in this secenario is due to  the vanishing of the  first term in the numerator of the $\log$-term (corresponding to  $\text{Re}\Pi_{c}$, Eq.~\ref{eq:RePi_c})  at the frequency $\omega_U$.  Once again we can pin-point the specific integral causing the divergence and it is due to 
\begin{widetext}
\begin{eqnarray}
I\propto \int \frac{pdp}{8\pi}\left[\left(1-\frac{p^2+\Delta _1 \Delta_2}{\sqrt{p^2+\Delta _1^2} \, \sqrt{p^2+\Delta _2^2}}\right)\frac{\Theta
\left(\sqrt{\mu ^2-\Delta _2^2}-p\right)}{w -\sqrt{p^2+\Delta _1^2}-\sqrt{p^2+\Delta_2^2}}\right],
\end{eqnarray}
\end{widetext}
where the contributions again arise from  $\Delta_1\rightarrow\Delta_2$ transition but now $\Delta_1$ and $\Delta_2$ corresponds to the lower and upper bands respectively.

\begin{figure}
\centering
\includegraphics[width=1.0\linewidth]{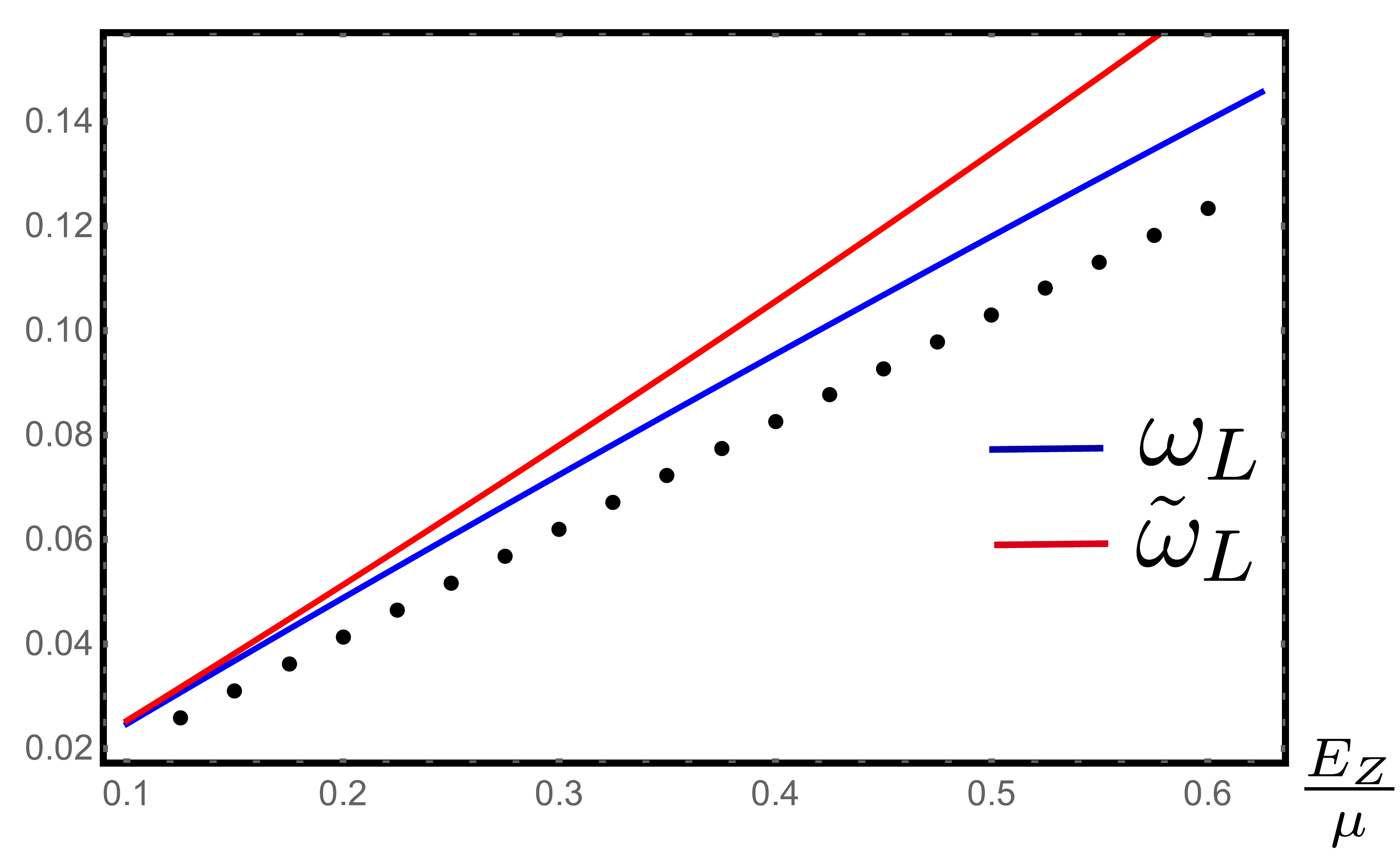}
\caption{(Color online) The dashed-dotted line denotes the pole position as a function of perpendicular electric field $E_Z$ for fixed screened interaction $u^*=0.5$, while blue and red solid curve corresponds to the behavior of two lower threshold frequency $\omega_L$ and $\tilde{\omega}_L$ with electric field, where in between imaginary part of polarization function is non-zero. }
\label{Pole_analysis}
\end{figure}

Solving  the pole equations  yield two solutions close to the threshold frequencies (see Fig.~\ref{Pole_position} for solution near the lower threshold) given by
\begin{eqnarray}
\omega_1&=&\omega_L-\frac{2\mu(\mu'-\mu)(\mu'-\mu''-2\mu)(\mu'+\mu''-2\mu)}{\mu'(\mu'-\mu'')(\mu'+\mu'')}\nonumber\\
&&\times \exp\big[{\frac{8\pi(\mu'-\mu )^3}{u^*\left[(\mu'-\mu )^2-\Delta_s^2\right]\Delta_d}}\big],
\end{eqnarray}
and a solution just below the upper threshold,
\begin{eqnarray}
\omega_2&=&\omega_U-\frac{2\mu(\mu+\mu'')(\mu'-\mu''-2\mu)(\mu'+\mu''+2\mu)}{\mu''(\mu'-\mu'')(\mu'+\mu'')}{}\nonumber\\
&& \times \exp\Big[{\frac{8\pi(\mu'-\mu )^3}{u^*\left[(\mu'-\mu )^2-\Delta_s^2\right]\Delta_d}}\Big],
\end{eqnarray}
where $\mu'=\sqrt{\mu^2 + \Delta_d\Delta_s}$  and $\mu''=\sqrt{\mu^2 - \Delta_d\Delta_s}$.  We note that in the absence of external electric field the two gaps  $\Delta_1$ and $\Delta_2$ are identical, therefore the Re$\Pi^0(\omega)$ vanishes identically and no pole solutions are possible.   In Fig.~(\ref{Pole_analysis}) we show the  explicit dependence of the threshold frequencies  $\omega_L$ and $\tilde{\omega}_L$   and the  lower pole  position on 
	the perpendicular electric field $E_Z$. For non-zero electric field, 
the slope of 	$\omega_L$ and $\tilde{\omega}_L$ are $2E_Z/\sqrt{\mu^2\pm4E_Z\lambda_{SO}},$ respectively, therefore the width of the real region given by $\tilde{\omega}_L-\omega_L$ grows wider. At the same time   the slope of pole position for a fixed screened interaction $u^*$ is even lesser than the slope of $\omega_L$ therefore the width between  the pole position and $\omega_L$ also increases.


While the above discussion hints at the possibility of collective excitations it turns out that the presence of the  sub-lattice degrees of freedom complicates the analysis. The pole equation has its structure modified due to the presence of 
$\tau_i$ type of terms in the Green's function. Even though $\Pi^{0}_{xx}$
has only $\sigma_x$ on either ends of the polarization bubble, the vertex corrected (due to electron-electron interactions) spin-susceptibility acquires contributions from all $\tau_i$'s.  For example, the lowest order vertex term  $\propto u\int \hat{G}_{P+Q}\sigma_x \hat{G}_{P}$ has $\tau$ dependence arising due to the Green's function. In terms of the  vertex term $\Lambda$ the interacting susceptibility, $\Pi_{xx}$, is given by 
\begin{eqnarray}
\Pi_{xx} = \int_P \text{Tr} [\sigma_x \hat{G}_P  \Lambda_x^0\hat{G}_{P+Q}],
\end{eqnarray}
where $\Lambda$ satisfies the  equation:
\begin{eqnarray}
\Lambda^\beta_j =\sigma_j \tau_\beta - u \int \hat{G}_{P}\Lambda^\beta_j \hat{G}_{P+Q}.\label{eq:vertex}
\end{eqnarray}
Under  the  assumption of momentum independent screened potential, $\Lambda$ will be a function of $Q$ only and is expressed as a linear combination of $\sigma_k \tau_\gamma $ (where $k,\gamma=0\cdot\cdot~ 3$)~\cite{maiti2015}. 
We express   $\Lambda^\beta_j$ as $\Lambda^\beta_j=  \sigma_k \tau_\gamma   M^{[4k + \gamma]}_{~~[4 j+\beta]}  $ (where 
the $16\times 16$ matrix $ M $ is a function of $Q$) in  Eq.~\ref{eq:vertex} and obtain
\begin{eqnarray}
(\sigma_k \tau_\gamma    +  u \int \hat{G}_{P}\sigma_k \tau_\gamma  \hat{G}_{P+Q} )  M^{[4k + \gamma]}_{~~[4 j+\beta]} =\sigma_j \tau_\beta.\label{eq:vertex2}
\end{eqnarray}
Multiplying both sides of Eq.~\ref{eq:vertex2} with $\sigma_m \tau_\nu $ and taking the trace yields,
\begin{eqnarray}
(\delta_{m,k} \delta_{\nu,\gamma} + \frac{ u}{4} \tilde{\Pi}^{[4m+\nu]}_{~~[4k+\gamma]}  ) M^{[4k + \gamma]}_{~~[4 j+\beta]} =\delta_{m,j} \delta_{\nu,\beta},\label{eq:vertex3}
\end{eqnarray}
where $\tilde{\Pi}$ is the generalized susceptibility whose elements are defined as $\tilde{\Pi}^{[4m+\nu]}_{~~[4j+\beta]}=\text{Tr}[\int_P  \sigma_m \tau_\nu   \hat{G}_{P}\sigma_j \tau_\beta \hat{G}_{P+Q}]$. The matrix M is thus given by  $M= (I+u\tilde{\Pi}/4)^{-1}$.  It turns out that many of the elements of $\tilde{\Pi}(\omega)$ matrix  exhibit  ultra-violet divergence. We will illustrate  one such example, consider the $\tilde{\Pi}_{55}$ element  given by $\tilde{\Pi}_{55}(\omega) = \text{Tr}[\int_P  \sigma_1 \tau_1   \hat{G}_{P}\sigma_1 \tau_1 \hat{G}_{P+Q}]$.  Here the terms independent of the chemical potential, i.e.,
\begin{eqnarray}
I_{\pm}\propto \int pdp \big(1+\frac{\Delta_1 \Delta_2}{ E_{p_1}E_{p_{2}}}\big) 
\frac{1}{  E_{p_1}+ E_{p_2}\pm\omega }  ,\label{eq:tildePF}
\end{eqnarray}
obtain divergent contributions from the upper limit due to the Dirac spectrum and  necessitates one to consider non-linear terms  arising from  the exact energy spectrum. The divergence of Eq.~\ref{eq:RePi} is expected to be altered in the interacting version $\Pi_{xx} = \int_P \text{Tr} [\sigma_x \hat{G}_P  \Lambda_x^0\hat{G}_{P+Q}]$, however, the  fate of collective excitations is not apriori clear, i.e., whether it   survives at all or survives with its peak position and peak width renormalized.

\vspace{0.5cm}
\subsection{Static limit: $\omega=0$}
Following earlier discusion,  the  components of spin-susceptibility that yield non-vanishing contributions are $\text{Re}\Pi_{zz}$ and $\text{Re}\Pi_{xx/yy}$. $\text{Re}\Pi_{zz}$ can be conveniently decomposed into the sum of $\text{Re}\Pi_{zz-1}+ \text{Re}\Pi_{zz-2} $ which are the contributions from transitions involving  $\Delta_{i}\rightarrow\Delta_{i}$($i=1,2$).
For $q<2k_{F_i}$,  $\text{Re}\Pi_{zz-i}$ is a constant. Subtracting the constant part we obtain 
\begin{widetext}
\begin{eqnarray}
\delta\text{Re}\Pi_{zz-i}=\left[\frac{\mu \sqrt{q^2-(2k_{F_{i}})^2}}{4\pi q}-\frac{\left(q^2-4 \Delta_i ^2\right)}{8\pi q}\tan^{-1}\left(\frac{\sqrt{q^2-(2k_{F_{i}})^2}}{2 \mu }\right)\right]\Theta(q-2k_{F_i}).
\end{eqnarray}
\end{widetext}
The above expression is identical to the charge susceptibility case~\cite{pyat2009, andreas2012,nicol2014,peeters2014,anmol2016}. For large distances the $zz$ component of the spin-susceptibility is given by 
\begin{eqnarray}
\chi_{zz-i}(r) \sim \int \text{d}q \sqrt{q}\frac{\cos(r q -\pi/4)}{\sqrt{r}}  \delta\Pi_{zz-i}(q).
\end{eqnarray}
Taking into  consideration that the first derivative of $\text{Re}\Pi_{zz-i}$ diverges at $2k_{F_i}$ as 
\begin{eqnarray}
\text{Re}\delta\Pi_{zz-i}'\approx \frac{\Delta_i^2}{\pi\mu\sqrt{2 k_{F_i}}}\frac{1}{\sqrt{q-2 k_{F_i}}}, 
\end{eqnarray}
the integral reduces to
\begin{eqnarray}
\chi_i(r) \sim \int  dq  \frac{\sqrt{q}\sin(r q -\pi/4)}{r^{3/2}   \sqrt{q-2k_{F_i}}}.
\end{eqnarray}  
Thus one can  deduce from simple power counting arguments  that at large distances  the $zz$ component of the  spin-susceptibility decays as $1/r^{2}$ and the contribution to exchange interaction is oscillatory with two wavelengths given by $\pi/k_{F_1}$ and $\pi/k_{F_2}$. 
For  electric field strength equal to  $E^{c}_z =\lambda_{SO}/l$, $\Delta_1=0$ and  therefore the first derivative of  $\text{Re}\Pi_{zz-1}$  vanishes.
It is the second derivative which diverges at $2k_{F_1}$ as  $\text{Re}\Pi_{zz-1}''\approx -(1/8\pi\sqrt{k_{F_i}})/\sqrt{q-2 k_{F_i}}$ that determines  the long distance behavior of $\text{Re}\chi_{zz-1}$. 
The susceptibility now acquires a faster  $1/r^3$ decay.
For $\mu>\Delta_2$  this behavior will be  masked by the slower $1/r^2$ decay arising due to $\text{Re}\chi_{zz-2}$, however for  $\mu<\Delta_2$,  only the $1/r^3$ term will survive.

Next consider the behavior of $\text{Re}\Pi_{xx/yy}$ (details of the derivation are given in the appendix~\ref{appen_B}). The terms which are independent of the chemical potential yield regular contributions for all values of $q$ given by
\begin{widetext}
\begin{eqnarray}
{\text{Re}}\,\Pi_a(q)=-\frac{\Delta _d^2+q^2 }{4 \pi  q^3}\Bigg\{\left[q^2-\Delta_s^2\right]\tan^{-1}\left(\frac{q}{\Delta _s}\right)+q\Delta_s\Bigg\}. 
\end{eqnarray}
While from the integrals containing $n_F(\sqrt{p^2+\Delta_1^2}-\mu)$ we obtain
\begin{eqnarray}
{\text{Re}}\Pi_b(q)=
\left\{  \begin{array}{lr}
-\frac{\mu-\Delta_1}{2\pi}-\frac{\text{sgn}\left(q^2+\Delta _2^2-\Delta _1^2\right)}{4\pi q}\big[Y(\mu)-Y(\Delta_1)\big],\quad{\text{for}}~~\,\, q<k_{F_1}-k_{F_2} \quad \text{or}\quad q>k_{F_1}+k_{F_2}, \\
\\
-\frac{\mu-\Delta_1}{2\pi}-\frac{\text{sgn}\left(q^2+\Delta _2^2-\Delta _1^2\right)}{4\pi q}\big[Y(\xi)-Y(\Delta_1)\big],\quad{\text{for}}~~\,\, k_{F_1}-k_{F_2}<q<k_{F_1}+k_{F_2},\label{eq:RePibq}
\end{array}
\right.
\end{eqnarray}
 where
\begin{eqnarray}
Y(x)= \left\{-2x\sqrt{\xi^2-x^2}-\tan^{-1}\left(\frac{x}{\sqrt{\xi^2-x^2}}\right)\Big[q^2-2\,\xi^2+\Delta_d^2\Big]\right\}.\label{eq:G_x}
\end{eqnarray}
\end{widetext}

\begin{figure}
\centering
\includegraphics[width=1.0\linewidth]{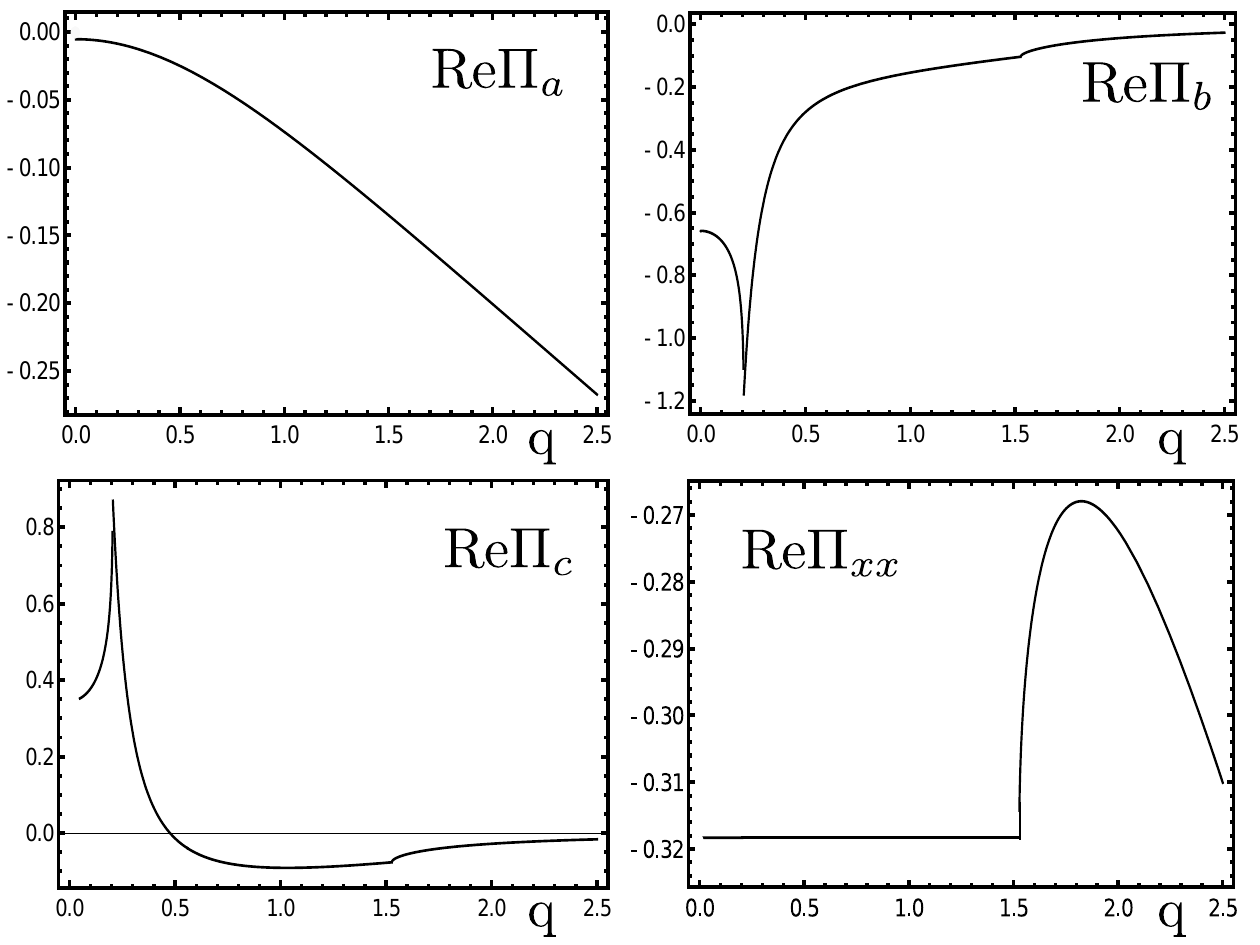}
\caption{(Color online) 
Here we consider the contributions to $\text{Re}\Pi_{xx/yy}(q,\omega=0)$. 
As in the appendix~\ref{appen_B}, we split the full integral  into  $\text{Re}\Pi_a$, $\text{Re}\Pi_b$ and  $\text{Re}\Pi_c$ and examine their behavior. The  contribution from $\text{Re}\Pi_a$ is smooth and continous. The sharp features of $\text{Re}\Pi_b$ and $\text{Re}\Pi_c$ at $k_{F_1}-k_{F_2}$ come with opposite sign, however, the kink like features at $k_{F_1}+k_{F_2}$ have same sign and when all three contributions are combined the features at $k_{F_1}+k_{F_2}$  are enhanced while those at $k_{F_1}-k_{F_2}$ cancel exactly.}
\label{RePi_w_0}
\end{figure}

The remaining term  arising from the integrals containing $n_F(\sqrt{p^2+\Delta_2^2}-\mu)$ denoted by $\text{Re}\Pi_c (q)$  is obtained by simply changing  $\Delta_1$ to $\Delta_2$ and vice-versa in  Eq.~\ref{eq:RePibq}.
The  derivatives of both  ${\text{Re}}\Pi_b(q)$ and ${\text{Re}}\Pi_c(q)$ diverge at $q_d=k_{F_1}-k_{F_2}$
and $q_s=k_{F_1}+k_{F_2}$ (see Fig.~\ref{RePi_w_0}).  However, combining them together we find that the  divergence at $q_d$ is cancelled and that $\text{Re}\Pi_{xx/yy}$ is constant for $q<q_s$, while the divergence at $q_s$ remains. Removing the constant part,  the full expression for the static-susceptibility is given by
\begin{widetext} 
\begin{eqnarray}
\delta\text{Re}\Pi_{xx/yy}(q)=\left[\frac{\mu \sqrt{\left(q^2-q_s^2\right) \left(q^2-q_d^2\right)}}{2\pi q^2} -\frac{\left(q^2+\Delta _d^2\right) \left(q^2-\Delta_s^2\right)}{4\pi q^3}\tan ^{-1}\left(\frac{\sqrt{\left(q^2-q_s^2\right) \left(q^2-q_d^2\right)}}{2\mu q}\right)\right]\Theta(q-k_{F_1}-k_{F_2}).\notag
\end{eqnarray}
\end{widetext}	
The derivative of the  polarization operator has a square-root singularity
at $q= q_s$ given  by,
\begin{eqnarray}
\delta\text{Re}\Pi_{xx/yy}'(q)\approx-\frac{\sqrt{k_{F_1}k_{F_2}} \big[\left(q^2+\Delta_d^2\right)
\left(q^2-\Delta_s^2\right)-4 \mu ^2 q_s^2\big]}{4 \sqrt{2} \pi  \mu 
q_s^{5/2} \sqrt{ q-q_s}},\notag
\end{eqnarray}
therefore the real space decay exhibits  $1/r^2$ power-law dependence at large distances while the  oscillatory wavelength is now given by $2\pi/q_s=2\pi/(k_{F_1}+k_{F_2})$. Rather interestingly, for the $xx$ and $yy$ parts of the spin-susceptibility, unless both the gaps are equal ($\Delta_1=\Delta_2$) closing of one of the gaps does not lead to  vanishing of the singular behavior of the derivative at $k_{F_1}+k_{F_2}$.
Thus the $1/r^2$ power-law dependence at large distances  is maintained irrespective of the tuning of the gaps by the electric field.

The real space analysis thus far yields the behavior of spin-spin correlation function between spins that are widely separated from each other and are delocalized on few sites. The calculation of spin-correlations thus entails disregarding intervalley scattering and taking the trace of the sub-lattice degrees of freedom. In contrast, the behavior of spin-correlations between two impurity spins that are localized on specific sites of the lattice is given by a different version of  static spin-susceptibility that also yields the  Rudermann-Kittel-Kasuya-Yosida (RKKY) interaction between the two localized spins (see~\cite{chang2015, duan2018} for a detailed analysis for the case of silicene). 
Due to the short-range nature of interactions between the localized impurities and itinerant electrons, an intervalley scattering of the electrons  via large $2K $ momentum exchange is allowed leading to additional contributions to the spin-susceptibility.  Moreover, the position of the spin-impurities (whether the two spins are on A-A/B-B sites or A-B sites) also crucially determines the behavior of spin-correlations. In what follows, we will briefly discuss the differences and similarities between  the results arising from these two different scenarios.

The effective interaction between two magnetic impurities $\vec{S}_{i}$ and $\vec{S}_{j}$ (localized at sites $\vec{R}_{i}$ and $\vec{R}_{j}$, respectively) is given by  
$H_{\text{RKKY}}=-J^2\chi^{cd}_{\alpha \beta}S_{i}^{\alpha}S_{j}^{\beta}$~\cite{kogan2011,jelena2013},
where there is a repeated summation on only  the spin indices $\alpha,\beta =x,y,z$; the indices  $c$, $d$  refer to the $A$ or $B$ sites and  $J$ is the interaction  term between the magnetic impurity and itinerant electrons.  
The spin-susceptibility matrix has the form,
\begin{eqnarray}
\chi_{\alpha,\beta}(R_{ij})=\frac{1}{\hbar}\int_{0}^{\infty} \text{Tr}[\sigma_\alpha  \mathcal{G}(i,j;\tau)\sigma_\beta  \mathcal{G}(j,i;-\tau)]d\tau,
\end{eqnarray}
where the trace is only over the spin degrees of freedom~\cite{kogan2011,jelena2013}.  The Green's function is a $4\times 4$ matrix,  
\begin{eqnarray}
\mathcal{G}(i,j;\pm \tau)=\mp\sum_{n}\psi_n(j)\psi^\dagger_n(i)e^{\mp\tilde{\epsilon}_n\tau}\Theta(\pm\tilde{\epsilon}_n),
\end{eqnarray}
where $n\in \{\eta, p,s\} $ is a summation on  valley, momentum and spin degrees of freedom,  $\tilde{\epsilon}_n =\epsilon_n-\mu$  and the wave-functions in the  basis $\psi_n=(\psi_{A\uparrow}, \psi_{B\uparrow},\psi_{A\downarrow},  \psi_{B\downarrow})^T$ are given by, 
\begin{equation}
\psi_{(\eta,p,\uparrow)}=\frac{e^{i(\eta K+p)R_i}} {\sqrt{2\epsilon_{\eta\uparrow}(\epsilon_{\eta\uparrow}+\nu\Delta_{\eta\uparrow})}}\begin{bmatrix}
pe^{-i\eta\theta} \\
\nu \epsilon_{\eta\uparrow}+\Delta_{\eta\uparrow} \\
0\\
0\\
\end{bmatrix} ,
\label{eq:scket}
\end{equation}
and 
\begin{equation}
\psi_{(\eta,p,\downarrow)}=\frac{e^{i(\eta K+p)R_i}} {\sqrt{2\epsilon_{\eta\downarrow}(\epsilon_{\eta\downarrow}+\nu\Delta_{\eta\downarrow})}}\begin{bmatrix}
0\\
0\\
pe^{-i\eta\theta} \\
\nu \epsilon_{\eta\downarrow}+\Delta_{\eta\downarrow} \\
\end{bmatrix},
\label{eq:scket}
\end{equation}
where $\nu=\pm 1$ represents conduction/valence band respectively. 

Let us for example consider $\chi^{\text{AA}}_{xx}$ and $\chi^{\text{AA}}_{xy}$ which are obtained from the following integrals
\begin{eqnarray}
\chi^{ \text{AA}}_{xx} =\int_0^\infty (g^{ \text{AA}}_{\uparrow\uparrow}\bar{g}^{ \text{AA}}_{\downarrow\downarrow} +  g^{ \text{AA}}_{\downarrow\downarrow}\bar{g}^{    \text{AA}}_{\uparrow\uparrow}) d\tau/\hbar
\end{eqnarray}  
and 
\begin{eqnarray}
\chi^{ \text{AA}}_{xy} =-i\int_0^\infty (g^{ \text{AA}}_{\uparrow\uparrow}\bar{g}^{ \text{AA}}_{\downarrow\downarrow} -  g^{ \text{AA}}_{\downarrow\downarrow}\bar{g}^{    \text{AA}}_{\uparrow\uparrow}) d\tau/\hbar
\end{eqnarray}  
 where $g^{ \text{AA}}_{ss} =e^{\mu\tau}\sum_{\eta} e^{i\eta \vec{K}.\vec{R}_{ij}} \mathcal{A}_{\eta,s}$ and

\begin{eqnarray}
 \mathcal{A}_{\eta ,s}=-\frac{a^2}{4\pi}\int\frac{p^3dp \Theta(\epsilon_{\eta s} -\mu)} {\epsilon_{\eta s}(\epsilon_{\eta s}+\Delta_{\eta s})}J_0(p|\vec{R}_{ij}|) e^{-\epsilon_{\eta s}\tau}.\quad
\end{eqnarray}
 Similarly 
$\bar{g}^{\text{AA}}_{ss} =e^{-\mu\tau}\sum_{\eta} e^{-i\eta \vec{K}.\vec{R}_{ij}}\mathcal{\bar{A}}_{\eta,s}$, 
where 
\begin{eqnarray}
 \mathcal{\bar{A}}_{\eta,s}=\frac{a^2}{4\pi}\int\sum_{\nu}\frac{p^3dp \Theta(\mu-\nu\epsilon_{\eta s} )  } {\epsilon_{\eta s}(\epsilon_{\eta s}+\nu\Delta_{\eta s})}      J_0(p|\vec{R}_{ij}|)  e^{\nu \epsilon_{\eta s}\tau}.\notag
\end{eqnarray}
Taking the product
\begin{eqnarray}
g^{\text{AA}}_{\uparrow\uparrow}\bar{g}^{\text{AA}}_{\downarrow\downarrow} =\sum_{\eta,\eta'}e^{i(\eta-\eta')\vec{K}.\vec{R}_{ij}}\mathcal{A}_{\eta,\uparrow}\bar{\mathcal{A}}_{\eta',\downarrow},\label{eq:gg}
\end{eqnarray}
and 
\begin{eqnarray}
g^{\text{AA}}_{\downarrow\downarrow}\bar{g}^{\text{AA}}_{\uparrow\uparrow} =\sum_{\eta,\eta'}e^{i(\eta-\eta')\vec{K}.\vec{R}_{ij}}\mathcal{A}_{\eta,\downarrow}\bar{\mathcal{A}}_{\eta',\uparrow},
\end{eqnarray}
we identify that the contributions can be classified into  intra  ($\eta=\eta'$)  and inter-valley ($\eta=-\eta'$) terms.
While for  $\chi^{\text{AA}}_{xx}$ the intra terms  add-up,   they  cancel identically for $\chi^{\text{AA}}_{xy}$. Similar cancellation holds for $\chi^{\text{AB}}_{xy}$.
This result is consistent with our earlier result (which takes into consideration only the intra terms) regarding the vanishing of $\chi_{xy}$ term when  contributions from the  valleys are added together. However due to the inter-valley scattering processes, $\chi^{\text{AB}}_{xy}$ and $\chi^{\text{AA}}_{xy}$, obtain  additional non-vanishing contributions. Another important difference is that, besides the oscillatory dependence with wave-number $2\pi/(k_{F_1}+k_{F_2})$  due to the intravalley process, the intervalley processes yield additional oscillatory dependence on $\vec{R}$ arising from terms  of the  type $e^{i2\eta\vec{K}.\vec{R}_{ij}}\mathcal{A}_{\eta,\uparrow}\bar{\mathcal{A}}_{-\eta,\downarrow}$ (see Eq.~\ref{eq:gg}).\\

\section{ Summary} {\label{sec:V}}

To summarize, in this article, we have presented a  detailed study of   the
 spin-susceptibility for silicene, that can be generalized to other buckled
  honeycomb structured materials e.g., germanene and stanene which also
   exhibit an electric field tunable band gap. We find that while the  $xx$ and $yy$ components 
  of the spin-susceptibility are identical, the $zz$ component is different. 
  The $xx$ and $yy$ components  obtain contributions from only those electronic transitions for which the 
 spins are flipped, while  the  $zz$   component obtain contributions from spin conserving processes.   
Although the off-diagonal components of the spin-susceptibility, $0z$ and   $xy$, are non-zero in individual valleys,  adding the   contributions from the valleys  leads to cancellation. The study of  the  imaginary part of spin-susceptibility reveals    regions in the $(q,\omega)$ plane where the  single-particle excitations  are allowed. We find that the threshold behavior for the lower to upper-band transition is especially interesting since its behavior changes upon increasing the value of $q$. 
For  $q$  values smaller than $q^*$ the threshold behavior exhibits a square-root singularity in its derivative, whereas for $q> q^*$ the susceptibility acquires a finite jump.  We have investigated the role of electric field $E_Z$ in  extending the allowed regions for particle-hole transitions. Electric field is also responsible for yielding differing contributions for the $\Delta_1\rightarrow\Delta_2$ transtions as compared to those from the $\Delta_2\rightarrow\Delta_1$ transitions.
Moreover, the magnitude of the $xx/yy$ components and $zz$ component also differ due to non-zero electric field.

We have studied the real part of spin-susceptibility, with particular emphasis on the dynamic and static limits.
In the dynamic limit, we show that the real part of spin-susceptibility exhibits log-divergence. 
The origin of  divergence at low frequencies can be  traced  to the  $u_{\Delta_1}\rightarrow u_{\Delta_2}$ transitions, whereas  those at high  frequencies can be attributed to  $l_{\Delta_1}\rightarrow u_{\Delta_2}$ transitions.  We  explore the  significance of the divergence 
for  spin-collective excitations and the dependence of the excitations  on external electric field. 
The study of the static part of the spin-susceptibility reveals  Kohn-anomaly
at $k_{F_1} +k_{F_2}$ for the $xx/yy$ components of the spin-susceptibility, whereas for the  $zz$ component
the anomaly is present at $2k_{F_1}$ and $2k_{F_2}$.
Tuning the electric field effects the behavior of the singularity at $2k_{F_1}$. We have explored the consequence of the Kohn-anomaly 
on the long distance behavior of the spin-susceptibility.
%

\section{ACKNOWLEDGMENTS}
S.S. and S.G. would like to thank S. Dutta, A. Kumar and V. Zyuzin for useful discussions. S.G. is grateful to SERB for the support via the grant number EMR/2016/002646.

\begin{widetext}
\section{ appendix}{\label{appen}}

\subsection{ Derivation of $\text{Re}\mathbf{\,\Pi_{xx/yy}(q,\omega)}$}  {\label{appen_A}}
We will integrate the terms of  Eq.~\ref{eq:PF1} by first obtaining the contribution from   $\Delta_1\rightarrow \Delta_2$ transition  by taking   $\beta=+1$ and $\beta'=-1$ (for all possible values of $\alpha,\alpha'$ for the   K-valley). We divide  the real part of polarization operator $\Pi(q,w)$ as follows, 

\begin{eqnarray}
A_1=-\frac{1}{2}\int\frac{d^2p}{(2\pi)^2}\left(1+\frac{{\vec{p}_1 \cdot \big(\vec{p}+\vec{q} \big)}_{2} }{E_1(p)\cdot E_2(p+q)}\right)\left[\frac{n_F[E_1(p)]}{-E_1(p) + E_2(p+q)-\omega} - \frac{n_F[E_2(p+q)]}{-E_1(p) + E_2(p+q)-\omega} \right]\notag\\
A_2=-\frac{1}{2}\int\frac{d^2p}{(2\pi)^2}\left(1-\frac{{\vec{p}_1 \cdot \big(\vec{p}+\vec{q} \big)}_{2} }{E_1(p)\cdot E_2(p+q)}\right)\left[\frac{1}{+E_1(p) + E_2(p+q)-\omega} - \frac{n_F[E_2(p+q)]}{+E_1(p) + E_2(p+q)-\omega}\right]\notag\\
A_3=-\frac{1}{2}\int\frac{d^2p}{(2\pi)^2}\left(1-\frac{{\vec{p}_1 \cdot \big(\vec{p}+\vec{q} \big)}_{2} }{E_1(p)\cdot E_2(p+q)}\right)\left[\frac{n_F[E_1(p)]}{-E_1(p) - E_2(p+q)-\omega} - \frac{1}{-E_1(p) - E_2(p+q)-\omega}\right],\notag
\end{eqnarray}
where $F^{1,-1}_{xx/yy}=2,~ \vec{p}_{1/2}=p_x \hat{e}_1 +\eta p_y \hat{e}_2 + \Delta_{1/2} \hat{e}_3, \quad E_1(p) = \sqrt{p^2+\Delta_1^2},$ and $E_2(p) = \sqrt{p^2+\Delta_2^2}$.
	
The first term of $A_2$ and second term of $A_3$ yield terms that are independent of $\mu$, we combine them together and represent it as $\Pi_{12-a}$. $\text{Im}\Pi_{12-a}$ is  given by 
\begin{eqnarray}
&&{\text{Im}}\,\Pi_{12-a}(q,\omega)=-\frac{1}{16}\Theta\Big(\omega^2-q^2-\Delta_s^2\Big){\text{Y}}(q,\omega),
\end{eqnarray} 
where
\begin{eqnarray}
&&{\text{Y}}(q,\omega)=\frac{1}{\sqrt{\omega^2-q^2}}\Bigg\{\Big[q^2 + 2\Delta_d^2\Big]
+\Big[\frac{2q^2(\Delta_1^2+\Delta_2^2)-2(\Delta_s\Delta_d)^2}{\omega^2-q^2}\Big]-\Big[ \frac{3q^2(\Delta_s\Delta_d)^2}{(\omega^2-q^2)^2}\Big] \Bigg\}.
\end{eqnarray} 
We use  the above result  to calculate  $\text{Re}\Pi_{12-a}$ via the Kramers-Kronig relation:
\begin{eqnarray}
{\text{Re}}\,\Pi_{12-a}(q,\omega)&&=\frac{1}{\pi}{\text{P}}\int_{-\infty}^{\infty} d\omega'\frac{{\text{Im}}\,\Pi_{12-a}(q,\omega')}{(\omega'-\omega)}\, {\text{sgn}}(\omega') 
= -\,\, \frac{1}{16\pi} {\text{P}}\Bigg(\int_{\gamma}^{\infty} d\omega'\frac{{\text{Y}}(q,\omega')}{(\omega'-\omega)}-\int_{-\infty}^{-\gamma} d\omega'\frac{{\text{Y}}(q,\omega')}{(\omega'-\omega)}\Bigg)\notag\\
&&=-\frac{1}{16\pi}\Big(\Theta\big(q-\omega\big)f(q,\omega)+\Theta\big(\omega-q\big)g(q,\omega)\Big).
\end{eqnarray}
The first integral is performed with the aid of  the following variable change $\omega'$ to $x$, where they are related via $\omega' =q (1+x^2)/(1-x^2)$.
Similar transformation is used for the second integral.

For $q>\omega$, the result of the integration is  $f(q,\omega)$ which is expressed as a sum of three parts, $f(q,\omega)=f_1(q,\omega)+f_2(q,\omega)+f_3(q,\omega)$ (corresponding to the square brackets of $Y(q,\omega)$) and they are given by, 
\begin{eqnarray}
&&f_1(q,\omega)=\Big(q^2 + 2\Delta_d^2\Big)\Bigg\{\frac{2}{(q+\omega)}\Bigg[\frac{1}{\tilde{\beta_1}}\,\,\,\tan^{-1}\Big(\frac{x}{\tilde{\beta_1}}\Big)\Bigg]+ \Big(\omega \rightarrow -\omega\Big)\Bigg\}_{\tan[\gamma'/2]}^{1}\nonumber\\
&&f_2(q,\omega)=\Big[2q^2(\Delta_1^2+\Delta_2^2)-2(\Delta_s\Delta_d)^2\Big]\Bigg\{\frac{1}{2(q+\omega)q^2}\Bigg[x-\frac{1}{\tilde{\beta_1}^2 x}-\frac{\left(\tilde{\beta_1}^2+1\right)^2 }{\tilde{\beta_1} ^3}\tan^{-1}\left(\frac{x}{\tilde{\beta_1}}\right)\Bigg] + \Big(\omega \rightarrow -\omega\Big)\Bigg\}_{\tan[\gamma'/2]}^{1}\nonumber\\
&&f_3(q,\omega)=\Bigg\{\frac{-3(\Delta_s\Delta_d)^2}{8(q+\omega)q^2}\Bigg[\frac{x^3}{3}-x\left(\tilde{\beta_1}^2+4\right) +\frac{4\tilde{\beta_1}^2+1}{\tilde{\beta_1}^4 x}-\frac{1}{3\tilde{\beta_1}^2 x^3}+\frac{\left(\tilde{\beta_1}^2+1\right)^4 }{\tilde{\beta_1}^5}\tan^{-1}\left(\frac{x}{\tilde{\beta_1}}\right)\Bigg] +  \Big(\omega \rightarrow -\omega\Big)\Bigg\}_{\tan[\gamma'/2]}^{1}.\nonumber
\end{eqnarray}
While  for $\omega>q$ regions, the result is expressed in terms of $g(q,\omega)$, where as before it is  expressed as sum of three parts,  $g(q,\omega)=g_1(q,\omega)+g_2(q,\omega)+g_3(q,\omega)$, which are given by
\begin{eqnarray}
&&g_1(q,\omega)=\Big(q^2 + 2\Delta_d^2\Big)\Bigg\{\frac{-2}{(q+\omega)}\Bigg[\frac{1}{\tilde{\beta_2}}\,\ln\bigg(\frac{x+\tilde{\beta_2}}{|x-\tilde{\beta_2}|}\bigg)\Bigg]+ \Big(\omega \rightarrow -\omega\Big)\Bigg\}_{\tan[\gamma'/2]}^{1}\nonumber\\
&&g_2(q,\omega)=\Big[2q^2(\Delta_1^2+\Delta_2^2)-2(\Delta_s\Delta_d)^2\Big]\Bigg\{\frac{-1}{2(q+\omega)q^2}\Bigg[-x-\frac{1}{\tilde{\beta_2}^2 x}+\frac{\left(\tilde{\beta_2}^2-1\right)^2 }{\tilde{\beta_2} ^3}\ln\bigg(\frac{x+\tilde{\beta_2}}{|x-\tilde{\beta_2}|}\bigg)\Bigg] + \Big(\omega \rightarrow -\omega\Big)\Bigg\}_{\tan[\gamma'/2]}^{1}\nonumber\\
&&g_3(q,\omega)=\Bigg\{\frac{3(\Delta_s\Delta_d)^2}{8(q+\omega)q^2}\Bigg[-\frac{x^3}{3}-x\left(\tilde{\beta_2}^2-4\right) +\frac{4\tilde{\beta_2}^2-1}{\tilde{\beta_2}^4 x}-\frac{1}{3\tilde{\beta_2}^2 x^3}+\frac{\left(\tilde{\beta_2}^2-1\right)^4 }{\tilde{\beta_2}^5}\ln\bigg(\frac{x+\tilde{\beta_2}}{|x-\tilde{\beta_2}|}\bigg)\Bigg] +  \Big(\omega \rightarrow -\omega\Big)\Bigg\}_{\tan[\gamma'/2]}^{1},
\end{eqnarray}
where $\gamma= \sqrt{q^2+\Delta_s^2}$,   $\gamma'=\cos^{-1}[q/\gamma]$, $\tilde{\beta_1}^2=(q-\omega)/(q+\omega)$, $\tilde{\beta_2}^2=(\omega-q)/(q+\omega)$ and $(\omega \rightarrow -\omega)$ represents similar terms with sign of $\omega$ changed. \\
As a next step,  $n_F[E_1(p)]$ terms from $A_1$ and $A_3$ are combined together and labelled  as $\text{Re}\Pi_{12-b}$:
\begin{eqnarray}
{\text{Re}}\,\Pi_{12-b}
&&=-\int\frac{d^2p}{(2\pi)^2}n_F[E_1(p)]\Bigg\{\frac{E_1(p) +\omega} {\big[E_2(p+q)\big]^2-\big[E_1(p)+\omega\big]^2} 
+\left[\frac{{\vec{p}_1 \cdot \big(\vec{p}+\vec{q} \big)}_{2} }{E_1(p)}\right]\frac{1} {\big[E_2(p+q)\big]^2-\big[E_1(p)+\omega\big]^2} \Bigg\}\notag\\
&&=-\frac{1}{4\pi}\left\{\int_{\Delta_1}^{\mu} \frac{ dE_1}{\sqrt{\omega^2-q^2}}\left[\frac{\Big((2E_1+\omega)^2-q^2-\Delta_d^2\Big) {\text{sgn}}\Big[\alpha_b-E_1\Big]}{\sqrt{\Big(2E_1+\omega\gamma_b\Big)^2-q^2\gamma_b^2 +\frac{4q^2\Delta_1^2}{\omega^2-q^2}}}\right]+(\mu-\Delta_1)\right\}'\notag
\end{eqnarray}
where
$\gamma_b=\left(\frac{\omega^2-q^2-\Delta_s\Delta_d}{\omega^2-q^2}\right)$, $\alpha_b=\left(\frac{q^2+\Delta_s\Delta_d-\omega^2}{2\omega}\right)$ and we have used, $\int_{0}^{2\pi}d\phi/(a+b\cos\phi)=2\pi~ \text{Sgn}[a]/\sqrt{a^2-b^2}$ to perform the angular integration.
Due to the  sgn function the result of the integration depends on the value of $\alpha_b$ with respect to the upper and lower limits, we obtain:
\begin{eqnarray}
&&(i).\quad\alpha_b > \mu \quad \Rightarrow \quad{\text{Re}}\,\Pi_{12-b}=
-\frac{1}{4\pi}\Re\left[\frac{1}{\sqrt{\omega^2-q^2}}\Big\{F_b(2\mu+\omega\gamma_b)-F_b(2\Delta_1+\omega\gamma_b)\Big\}+(\mu-\Delta_1)\right]\notag\\
&&(ii).\quad\mu>\alpha_b> \Delta_1 \quad \Rightarrow \quad {\text{Re}}\,
\Pi_{12-b}=	-\frac{1}{4\pi}\Re\left[\frac{1}{\sqrt{\omega^2-q^2}}\Big\{F_b(2\mu+\omega\gamma_b)+
F_b(2\Delta_1+\omega\gamma_b)-2F_b(2\alpha_b+\omega\gamma_b)\Big\}\right]\notag\\
&&(iii).\quad\alpha_b < \Delta_1 \quad\Rightarrow \quad {\text{Re}}\,\Pi_{12-b}=
+\frac{1}{4\pi}\Re\left[\frac{1}{\sqrt{\omega^2-q^2}}\Big\{F_b(2\mu+\omega\gamma_b)-F_b(2\Delta_1+\omega\gamma_b)\Big\}+(\mu-\Delta_1)\right],\notag\\
\end{eqnarray}
where $\Re$ represents the real part of the corresponding function and 
\begin{eqnarray}
F_b(x) =\frac{1}{2} \left[\Big(\xi_b^2-2\Delta_d^2-2q^2+2\big(\omega\gamma_b-\omega\big)^2 \Big)\log\left(\sqrt{x^2-\xi_b^2}+x\right)+\Big(x-4\big(\omega\gamma_b-\omega\big)\Big) \sqrt{x^2-\xi_b^2}\right]\notag
\end{eqnarray}
and  $\xi_b=\sqrt{q^2\gamma_b^2 -\frac{4q^2\Delta_1^2}{\omega^2-q^2}}$.\\
	
Finally  the terms corresponding to $n_F[E_2(p+q)]$ from $A_1$ and $A_2$  are combined together into  Re$\,\Pi_{12-c}$:\\
\begin{eqnarray}
{\text{Re}}\,\Pi_{12-c}
&&=\int\frac{d^2p}{(2\pi)^2}n_F[E_2(p)]\Bigg\{\frac{\omega-E_2(p)} {\big[E_1(p+q)\big]^2-\big[E_2(p)-\omega\big]^2} 
-\left[\frac{{\big(\vec{p}+\vec{q} \big)}_{1}\cdot\vec{p}_2}{ E_2(p)}\right]\frac{1} {\big[E_1(p+q)\big]^2-\big[E_2(p)-\omega\big]^2} \Bigg\}\notag\\
&&=	-\frac{1}{4\pi}\left\{\int_{\Delta_2}^{\mu} \frac{ dE_2}{\sqrt{\omega^2-q^2}}\left[\frac{\Big((2E_2-\omega)^2-q^2-\Delta_d^2\Big) {\text{Sgn}}\Big[E_2-\alpha_c\Big]}{\sqrt{\Big(2E_2-\omega\gamma_c\Big)^2-q^2\gamma_c^2 +\frac{4q^2\Delta_2^2}{\omega^2-q^2}}}\right]+(\mu-\Delta_2)\right\},\notag
\end{eqnarray}
where	$\gamma_c=\left(\frac{\omega^2-q^2+\Delta_s\Delta_d}{\omega^2-q^2}\right)$ and $\alpha_c=\left(\frac{\omega^2-q^2+\Delta_s\Delta_d}{2\omega}\right)$.
As before, due to the  sgn function,  the integral yields  three different results  depending on the value of $\alpha_c$.  They are
\begin{eqnarray}
&&(i).\quad\alpha_c > \mu \quad \Rightarrow \quad{\text{Re}}\,\Pi_{12-c}=
+\frac{1}{4\pi}\Re\left[\frac{1}{\sqrt{\omega^2-q^2}}\Big\{F_c(2\mu-\omega\gamma_c)-F_c(2\Delta_2-\omega\gamma_c)\Big\}+(\mu-\Delta_2)\right]\notag\\
&&(ii).\quad\mu>\alpha_c> \Delta_2 \quad \Rightarrow \quad {\text{Re}}\,\Pi_{12-c}=
-\frac{1}{4\pi}\Re\left[\frac{1}{\sqrt{\omega^2-q^2}}\Big\{F_c(2\mu-\omega\gamma_c)+F_c(2\Delta_2-\omega\gamma_c)-2F_c(2\alpha_c-\omega\gamma_c)\Big\}\right]\notag\\
&&(iii).\quad\alpha_c < \Delta_2 \quad\Rightarrow \quad {\text{Re}}\,\Pi_{12-c}=
-\frac{1}{4\pi}\Re\left[\frac{1}{\sqrt{\omega^2-q^2}}\Big\{F_c(2\mu-\omega\gamma_c)-F_c(2\Delta_2-\omega\gamma_c)\Big\}+(\mu-\Delta_2)\right],\notag\\
\end{eqnarray}
where
\begin{eqnarray}
F_c(x) =\frac{1}{2} \left[\Big(\xi_c^2-2\Delta_d^2-2q^2+2\big(\omega\gamma_c-\omega\big)^2 \Big)\log\left(\sqrt{x^2-\xi_c^2}+x\right)+\Big(x+4\big(\omega\gamma_c-\omega\big)\Big) \sqrt{x^2-\xi_c^2}\right],\notag
\end{eqnarray}
and  $\xi_c=\sqrt{q^2\gamma_c^2 -\frac{4q^2\Delta_2^2}{\omega^2-q^2}}$.

Similar to the earlier derivation  we will next integrate the terms of Eq.~\ref{eq:PF1} by considering the contributions from   $\Delta_2\rightarrow \Delta_1$ transition  by considering $\beta=-1$ and $\beta'=+1$ (in the case of K-valley), for all possible values of $\alpha$ and $\alpha'$. As before, we divide  the real part of polarization operator $\Pi(q,w)$ as follows, 
\begin{eqnarray}
B_1=-\frac{1}{2}\int\frac{d^2p}{(2\pi)^2}\left(1+\frac{{\vec{p}_2 \cdot \big(\vec{p}+\vec{q}\big)}_1}{E_2(p)\cdot E_1(p+q)}\right)\left[\frac{n_F[E_2(p)]}{-E_2(p) + E_1(p+q)-\omega} - \frac{n_F[E_1(p+q)]}{-E_2(p) + E_1(p+q)-\omega} \right]\notag\\
B_2=-\frac{1}{2}\int\frac{d^2p}{(2\pi)^2}\left(1-\frac{{\vec{p}_2\cdot \big(\vec{p}+\vec{q}\big)}_1}{E_2(p)\cdot E_1(p+q)}\right)\left[\frac{1}{+E_2(p) + E_1(p+q)-\omega} - \frac{n_F[E_1(p+q)]}{+E_2(p) + E_1(p+q)-\omega}\right]\notag\\
B_3=-\frac{1}{2}\int\frac{d^2p}{(2\pi)^2}\left(1-\frac{{\vec{p}_2\cdot \big(\vec{p}+\vec{q}\big)}_1}{E_2(p)\cdot E_1(p+q)}\right)\left[\frac{n_F[E_2(p)]}{-E_2(p) - E_1(p+q)-\omega} - \frac{1}{-E_2(p) - E_1(p+q)-\omega}\right],\notag
\end{eqnarray}
		
The first term of $B_2$ and the second term of $B_3$ yield terms that are independent of $\mu$, we combine them together and represent it as Re$\,\Pi_{21-a}$. Performing the following change of variables  $p+q\rightarrow p$ and $p\rightarrow-p$ it is easy to show that  $\text{Re}\Pi_{21-a}(q,\omega)=\text{Re}\Pi_{12-a}(q,\omega)$. The combined contribution represented as $\text{Re}\Pi_{a}$ is thus given by  $\text{Re}\Pi_{a}=\text{Re}\Pi_{21-a}(q,\omega)+\text{Re}\Pi_{12-a}(q,\omega)$.
		
Similar to the evaluation of  Re$\,\Pi_{12-b}$, we combine  terms corresponding to $n_F[E_1(p+q)]$ from $B_1$ and $B_2$  and denote the contributions  as $\text{Re}\,\Pi_{21-b}$. Change of variables as above yields,
\begin{eqnarray}
{\text{Re}}\,\Pi_{21-b}(q,\omega)
&&=-\int\frac{d^2p}{(2\pi)^2}n_F[E_1(p)]\Bigg\{\frac{E_1(p) -\omega} {\big[E_2(p+q)\big]^2-\big[E_1(p)-\omega\big]^2} 
+\left[\frac{{\vec{p}_1 \cdot \big(\vec{p}+\vec{q} \big)}_{2} }{E_1(p)}\right]\frac{1} {\big[E_2(p+q)\big]^2-\big[E_1(p)-\omega\big]^2} \Bigg\},\notag
\end{eqnarray}
thus ${\text{Re}}\,\Pi_{21-b}(q,\omega)={\text{Re}}\,\Pi_{12-b}(q,-\omega)$. The total contribution is,  ${\text{Re}}\,\Pi_{b}(q,\omega)={\text{Re}}\,\Pi_{12-b}(q,\omega)+{\text{Re}}\,\Pi_{21-b} (q,\omega)$.
Following essentially same arguments we obtain
${\text{Re}}\,\Pi_{21-c}(q,\omega)={\text{Re}}\,\Pi_{12-c}(q,-\omega)$, thus ${\text{Re}}\,\Pi_{c}(q,\omega)={\text{Re}}\,\Pi_{12-c}(q,\omega)+{\text{Re}}\,\Pi_{21-c}(q,\omega)$.
Therefore, the full result for $\text{Re}\Pi_{xx/yy}$ is
\begin{eqnarray}
\text{Re}\Pi_{xx/yy} = \text{Re}\Pi_{a}(q,\omega) + \text{Re}\Pi_{b}(q,\omega)+ \text{Re}\Pi_{c}(q,\omega) .
\end{eqnarray}
		
\subsection{ $\text{Re}\mathbf{\,\Pi_{xx,yy}(q,\omega=0)}$}{\label{appen_B}}
We will use the expression of $\text{Re}\Pi_{xx,yy}(q,\omega)$ as given in Appendix \ref{appen_A} to obtain the $\omega=0$ limit. As before, $\text{Re}\Pi_{xx,yy}$ can be  expressed as sum of three components, $\text{Re}\Pi_{xx,yy}(q)$= $\text{Re}\Pi_a(q)$+ $\text{Re}\Pi_b(q)$+ $\text{Re}\Pi_c(q)$.
The results of the calculations for the individual terms are as follows. The integral without the chemical potential term is given by   	
\begin{eqnarray}
{\text{Re}}\,\Pi_a(q)=-\frac{1}{8\pi}\Big(f_1+f_2+f_3\Big),
\end{eqnarray}
where the lower limit on all three integrals are 	 $l=\tan\Bigg[\frac{1}{2}\cos^{-1}\Big(\frac{q}{\sqrt{q^2+\Delta_s^2}}\Big)\Bigg]$,
\begin{eqnarray}
f_1(q,0)&=&\frac{4}{q}\Big[q^2 + 2\Delta_d^2\Big]\Bigg[\tan^{-1}\big(x\big)\Bigg]^1_{l},\\
f_2(q,0)&=&\frac{1}{q^3}\Big[2q^2(\Delta_1^2+\Delta_2^2)-2(\Delta_s\Delta_d)^2\Big]\Bigg[x-\frac{1}{x}-4\tan^{-1}(x)\Bigg]^1_{l},\\\nonumber
f_3(q,0)&=&-\frac{3(\Delta_s\Delta_d)^2}{4q^3}\Bigg[\frac{x^3}{3}-5x+\frac{5}{x}-\frac{1}{3x^3}+2^4\tan^{-1}(x)\Bigg]^1_{l}.\nonumber 
\end{eqnarray}
Combining them together we obtain,
$
{\text{Re}}\,\Pi_a(q)=-\frac{\Delta_d^2+q^2 }{4 \pi  q^3}\Bigg\{\left[q^2-\Delta_s^2\right]\tan^{-1}\left(\frac{q}{\Delta _s}\right)+q\Delta_s\Bigg\}.$\\
$\text{Re}\Pi_b(q)$ includes contributions from all integrals that have  $n_F[E_1(p)]$ and $n_F[E_1(p+q)]$  terms:
\begin{eqnarray}
{\text{Re}}\,\Pi_b(q)=-\frac{1}{2\pi }\Bigg\{\mu-\Delta_1+  \int_{\Delta_1}^{\mu} \frac{dx}{2q}\frac{\Big[4x^2-q^2-\Delta_d^2\Big]{\text{sgn}} \Big(q^2+\Delta_s\Delta_d\Big)}{\sqrt{\xi^2-x^2}}\Bigg\},
\end{eqnarray}
where  $\xi=\sqrt{\frac{\big(q^2+\Delta_s\Delta_d\big)^2+4q^2\Delta_1^2}{4q^2}}$. The result of the integration is  given in  Eq.~\ref{eq:RePibq} of the main text.
The last term, $\text{Re}\Pi_c(q)$, includes  contributions from  all integrals containing  $n_F[E_2(p)]$ and $n_F[E_2(p+q)]$ terms and is given by
\begin{eqnarray}
{\text{Re}}\,\Pi_c(q)=-\frac{1}{2\pi }\Bigg\{\mu-\Delta_2+  \int_{\Delta_2}^{\mu} \frac{dx}{2q}\frac{\Big[4x^2-q^2-\Delta_d^2\Big]{\text{sgn}} \Big(q^2-\Delta_s\Delta_d\Big)}{\sqrt{\xi^2-x^2}}\Bigg\},
\end{eqnarray}
Final result for  ${\text{Re}}\,\Pi_c(q)$ is obtained from ${\text{Re}}\,\Pi_b(q)$ by exchanging $\Delta_1$ with $\Delta_2$ and vice-versa.   Adding together the three terms we find that the static part of the  polarization function has a constant value for  $q<k_{F_1}+k_{F_2}$, while the    change in polarization function from the constant value for $q>k_{F_1}+k_{F_2}$   is given by,
\begin{eqnarray}
\delta\left[ {\text{Re}}\Pi_{xx,yy}(q) \right]=\frac{1}{2\pi}\left[\frac{\mu\sqrt{(q^2-q_d^2)(q^2-q_s^2)}}{q^2}-\frac{(q^2+\Delta_d^2)(q^2-\Delta_s^2)}{2q^3}\tan ^{-1}\left(\frac{\sqrt{(q^2-q_d^2)(q^2-q_s^2)}}{2\mu q}\right)\right],
\end{eqnarray}
where $q_s=k_{F_1}+k_{F_2}$ and $q_d=k_{F_1}-k_{F_2}$.
			

\subsection{ Derivation of Re$\,\mathbf{\Pi_{xx,yy}(q=0,\omega)}$}{\label{appen_C}}
The Im$\,\Pi_a(\omega)$ term  with contributions from both  $\Delta_1\rightarrow\Delta_2$ and $\Delta_2\rightarrow\Delta_1$ transitions is given by,
\begin{eqnarray}
{\text{Im}}\,\Pi_a(\omega)=-\frac{1}{8}\Theta\Big(\omega^2-\Delta_s^2\Big){\text{Y}}(\omega);
\quad{\text{Y}}(\omega)=\frac{2\Delta_d^2}{\omega}\left[1
-\frac{\Delta_s^2}{\omega^2}\right].
\end{eqnarray} 
Utilizing the Kramers-Kronig relation we obtain for Re$\,\Pi_a(\omega)$
\begin{eqnarray}
{\text{Re}}\,\Pi_a(\omega)=\frac{1}{\pi}{\text{P}}\int_{-\infty}^{\infty} d\omega'\frac{{\text{Im}}\,\Pi_a(\omega')}{(\omega'-\omega)}\, {\text{sgn}}(\omega')
=-\frac{\Delta_d^2}{4\pi\omega}\left\{ \log\left[\frac{\Delta_s+\omega}{|\Delta_s-\omega|}\right] \left(1-\frac{\Delta_s^2}{\omega^2}\right)+\frac{2\Delta_s}{\omega}\right\}.\notag
\end{eqnarray}
A direct integration by considering contributions from the integrals containing $n_F(E_1)$ term yields				
\begin{eqnarray}
{\text{Re}}\,\Pi_b
&&=-\int\frac{d^2p}{(2\pi)^2}n_F[E_1(p)]\Bigg\{\frac{E_1(p)+\omega} {\big[E_2(p)\big]^2-\big[E_1(p)+\omega\big]^2} 
+\left[\frac{{\vec{p}_1\cdot\vec{p}}_{2} }{E_1(p)}\right]\frac{1} {\big[E_2(p)\big]^2-\big[E_1(p)+\omega\big]^2} \Bigg\}+\Big[\omega\rightarrow-\omega\Big]\notag\\
&&=-\frac{1}{4\pi}\left\{\int_{\Delta_1}^{\mu} dE_1
\left[\frac{(2E_1+\omega)^2-\Delta_d^2}{\Delta_s\Delta_d-\omega^2-2E_1\omega}\right]+\Big[\omega\rightarrow-\omega\Big]\right\}-\frac{(\mu-\Delta_1)}{2\pi}\notag\\
&&=-\frac{1}{4\pi} \Bigg\{ \frac{\Delta_d^2 \left(\Delta_s^2-\omega ^2\right)}{2\omega^3}\left( \log \left[\frac{\left(-\Delta_d\Delta_s-2\mu\omega+\omega ^2\right)   \left(-\Delta_d\Delta_s+2\omega\Delta_1+\omega ^2\right)}{\left(-\Delta_d\Delta_s+2\mu\omega+\omega ^2\right)   \left(-\Delta_d\Delta_s-2\omega\Delta_1+\omega ^2\right)}\right] \right)-\frac{2\Delta_d\Delta_s(\mu-\Delta_1)}{\omega^2}\Bigg\}.
\end{eqnarray}	
Similarly, $n_F(E_2)$ term yields	contribution to ${\text{Re}}\,\Pi_c$ given by
\begin{eqnarray}
{\text{Re}}\,\Pi_c
&&=-\int\frac{d^2p}{(2\pi)^2}n_F[E_2(p)]\Bigg\{\frac{E_2(p)-\omega} {\big[E_1(p)\big]^2-\big[E_2(p)-\omega\big]^2} 
+\left[\frac{\vec{p}_{1}\cdot\vec{p}_2}{ E_2(p)}\right]\frac{1} {\big[E_1(p)\big]^2-\big[E_2(p)-\omega\big]^2} \Bigg\}+\Big[\omega\rightarrow-\omega\Big]\notag\\
&&=-\frac{1}{4\pi}\left\{\int_{\Delta_2}^{\mu} dE_2
\left[\frac{(2E_2-\omega)^2-\Delta_d^2}{-\Delta_s\Delta_d-\omega^2+2E_2\omega}\right]+\Big[\omega\rightarrow-\omega\Big]\right\}-\frac{(\mu-\Delta_2)}{2\pi}\notag\\
&&=-\frac{1}{4\pi} \Bigg\{ \frac{\Delta_d^2 \left(\Delta_s^2-\omega ^2\right)}{2\omega^3}\left( \log \left[\frac{\left(\Delta_d\Delta_s-2\mu\omega+\omega ^2\right)\left(\Delta_d\Delta_s+2\omega\Delta_2+\omega ^2\right)}{\left(\Delta_d\Delta_s+2\mu\omega+\omega ^2\right)\left(\Delta_d\Delta_s-2\omega\Delta_2+\omega ^2\right)}\right] \right)+\frac{2\Delta_d\Delta_s(\mu-\Delta_2)}{\omega^2}\Bigg\}.\notag\\
\end{eqnarray}

\end{widetext}

\bibliography{Silicene_ref} 

\end{document}